 \documentclass[sigconf, screen]{acmart}
\usepackage{array}
\newcolumntype{P}[1]{>{\centering\arraybackslash}p{#1}}

\usepackage{listings}
\usepackage{xcolor} 
\usepackage{multirow}
\usepackage{booktabs} 
\usepackage{amsmath}
\usepackage{graphicx}
\usepackage[belowskip=-12pt]{caption}
\usepackage{subcaption}
\usepackage{rotating}
\usepackage{tabularx}
\usepackage[noend,linesnumbered, ruled]{algorithm2e}
\usepackage{wrapfig}
\usepackage{listings}
\usepackage{color, colortbl}
\usepackage{balance} 
\usepackage{lscape}
\SetKw{KwStep}{Step}
\usepackage{hyperref}
\usepackage{xspace}
\usepackage{framed}
\usepackage{syntax}
\usepackage{comment}
\usepackage[flushleft]{threeparttable}
\newcolumntype{C}[1]{>{\centering\arraybackslash}p{#1}}

\usepackage{caption} 
\usepackage{caption} 
\usepackage[framemethod=TikZ]{mdframed}
\usepackage{soul}
\usepackage{flushend}
\usepackage{enumitem}

\usepackage{tikz}

\usepackage{natbib}

\definecolor{codegreen}{rgb}{0,0.6,0}
\definecolor{codegray}{rgb}{0.5,0.5,0.5}
\definecolor{codepurple}{rgb}{0.58,0,0.82}
\definecolor{backcolour}{rgb}{0.95,0.95,0.92}
\definecolor{Gray}{gray}{0.1}
\lstdefinestyle{mystyle}{
	backgroundcolor=\color{backcolour},   
	commentstyle=\color{codegreen},
	keywordstyle=\color{magenta},
	numberstyle=\tiny\color{codegray},
	stringstyle=\color{codepurple},
	basicstyle=\scriptsize,
	breakatwhitespace=false,         
	breaklines=true,                 
	captionpos=b,                    
	keepspaces=true,                 
	numbers=left,                    
	numbersep=5pt,                  
	showspaces=false,                
	showstringspaces=false,
	showtabs=false,                  
	tabsize=2
}

\lstdefinelanguage{Pythonna}{%
	language     = python,
	morekeywords = {to_categorical, flow_from_directory, pad_sequences, load_image}
}

\lstdefinestyle{customc}{
	belowcaptionskip=1\baselineskip,
	breaklines=false,
	frame= single,
	breaklines = true,
	xleftmargin=\parindent,
	language= Pythonna,
	showstringspaces=false,
	basicstyle=\footnotesize\ttfamily,
	keywordstyle=\bfseries\color{green!40!black},
	commentstyle=\itshape\color{purple!40!black},
	identifierstyle=\color{blue},
	stringstyle=\color{codegreen},
	backgroundcolor=\color{gray!4}
}

\graphicspath{{./figures/}}

\newcommand{\fignref}[1]{Figure~\ref{#1}}
\newcommand{\lstref}[1]{Listing~\ref{#1}}

\newcommand{\secref}[1]{\S\ref{#1}}

\newcommand{\etal}{{\em et al.}\xspace}

\newcommand{\sof}{\textit{Stack Overflow}\xspace}
\newcommand{\keras}{\textit{Keras}\xspace}

\newcommand{\gh}{GitHub\xspace}

\newcounter{rqs}
\stepcounter{rqs}

\newcounter{NumObservations}
\stepcounter{NumObservations}
\definecolor{shadecolor}{rgb}{.9,.9,.9}

\AtBeginDocument{%
  \providecommand\BibTeX{{%
    \normalfont B\kern-0.5em{\scshape i\kern-0.25em b}\kern-0.8em\TeX}}}

\setcopyright{none} 
\copyrightyear{2022}
\acmYear{2022}
\acmDOI{} 

\acmConference[ICSE 2022]{The 44th International Conference on Software Engineering}{May 21–29, 2022}{Pittsburgh, PA, USA}

\acmPrice{}
\acmISBN{978-1-4503-XXXX-X/18/06}

\begin{document}

\title{DeepDiagnosis: Automatically Diagnosing Faults and Recommending Actionable Fixes in Deep Learning Programs}

\subtitle{}



\author{Mohammad Wardat}
\email{wardat@iastate.edu}
\affiliation{%
	\institution{Dept. of Computer Science, Iowa State University}
	\streetaddress{226 Atanasoff Hall}
	\city{226 Atanasoff Hall, Ames}
	\state{IA}
	\postcode{50010}
	\country{USA}
}

\author{Breno Dantas Cruz}
\email{bdantasc@iastate.edu}
\affiliation{%
	\institution{Dept. of Computer Science, Iowa State University}
	\streetaddress{226 Atanasoff Hall}
	\city{226 Atanasoff Hall, Ames}
	\state{IA}
	\postcode{50010}
	\country{USA}
}

\author{Wei Le}
\email{weile@iastate.edu}
\affiliation{%
	\institution{Dept. of Computer Science, Iowa State University}
	\streetaddress{226 Atanasoff Hall}
	\city{226 Atanasoff Hall, Ames}
	\state{IA}
	\postcode{50010}
	\country{USA}
}

\author{Hridesh Rajan}
\email{hridesh@iastate.edu}
\affiliation{%
	\institution{Dept. of Computer Science, Iowa State University}
	\streetaddress{226 Atanasoff Hall}
	\city{226 Atanasoff Hall, Ames}
	\state{IA}
	\postcode{50010}
	\country{USA}
}




\begin{abstract}
Deep Neural Networks (DNNs) are used in a wide variety of applications. However, as in any software application, DNN-based apps are afflicted with bugs. Previous work observed that DNN bug fix patterns are different from traditional bug fix patterns. Furthermore, those buggy models are non-trivial to diagnose and fix due to inexplicit errors with several options to fix them. To support developers in locating and fixing bugs, we propose DeepDiagnosis, a novel debugging approach that localizes the faults, reports error symptoms and suggests fixes for DNN programs. In the first phase, our technique monitors a training model, periodically checking for eight types of error conditions. Then, in case of problems, it reports messages containing sufficient information to perform actionable repairs to the model. In the evaluation, we thoroughly examine 444 models -- 53 real-world from \gh and \sof, and 391 curated by AUTOTRAINER. DeepDiagnosis provides superior accuracy when compared to UMLUAT and DeepLocalize. Our technique is faster than AUTOTRAINER for fault localization. The results show that our approach can support additional types of models, while state-of-the-art was only able to handle classification ones. Our technique was able to report bugs that do not manifest as numerical errors during training. Also, it can provide actionable insights for fix whereas DeepLocalize can only report faults that lead to numerical errors during training. DeepDiagnosis manifests the best capabilities of fault detection, bug localization, and symptoms identification when compared to other approaches. 

\end{abstract}





\maketitle

\section{Introduction}
\label{sec:intro}

Deep Neural Networks (DNNs) are becoming increasingly popular due to their successful applications in many areas, such as healthcare~\cite{janowczyk2016deep, miotto2018deep}, transportation~\cite{veres2019deep}, and entertainment~\cite{grossman2009survey}. But, the intrinsic complexity of deep learning apps requires that developers build DNNs within their software systems to facilitate integration and development with other applications. The construction of such systems requires popular Deep Learning libraries~\cite{TensorFlow,Keras}.

Despite the increasing popularity and many successes for using Deep Learning libraries and frameworks, DNN applications still suffer from reliability issues~\cite{zhang2006locating,islam20repairing,islam19}. These faults are harder to detect and debug when compared to traditional software systems, as the bugs are often obfuscated within the DNNs. Therefore, it is important and necessary to diagnose their faults, and provide actionable fixes. To that end, software engineering research has recently focused on improving the reliability of DNN-based software. For instance, there have been studies on characterizing DNN bugs~\cite{zhang2006locating,islam20repairing,islam19}, on testing frameworks for deep 
learning~\cite{DeepTest}, on debugging deep learning using differential
analysis~\cite{ma2018mode}, and fixing DNNs~\cite{zhang2019apricot,ZhangAUTOTRAINER,DeepFault}.
There are also frameworks and tools for inspecting and detecting unexpected behavior in DNNs. However, they require that specialists verify the visualization, which is only available upon completing the training phase~\cite{mane2015tensorboard,tfdbg,Visdom,TensorWatch,Manifold}.

Due to the complexity of using existing frameworks to debug and localize 
faults in deep learning software, recent SE research has introduced techniques for automatically localizing bugs~\cite{wardat21DeepLocalize,zhang2020detecting}. DeepLocalize performs dynamic analysis during training to localize bugs by monitoring values produced at the intermediate nodes of the DNNs~\cite{wardat21DeepLocalize}. If there is a numerical error, then this approach traces that back to the faulty layer. DEBAR~\cite{zhang2020detecting} is a static analysis tool that detects numerical errors in the DNNs. While both approaches have significantly advanced the state of the art in debugging DNNs, they do not detect bugs that manifest as trends of values (e.g. vanishing gradient, exploding gradient, accuracy not increasing) and do not offer possible fixes.




We propose DeepDiagnosis (DD), an approach for localizing faults, reporting error symptoms, diagnosing problems, and providing suggestions to fix structural bugs in DNNs. Our approach introduces three new symptoms of structural bugs and defines new rules to map fault location to its root cause in DNN programs. We implemented DD as a dynamic analysis tool and compared and contrasted it against state-of-the-art approaches. DD outperforms UMLAUT~\cite{schoop2021umlaut} and DeepLocalize~\cite{wardat21DeepLocalize} in terms of efficiency and AUTOTRAINER in terms of performance~\cite{ZhangAUTOTRAINER}.
For example, assume the \textit{unchanged weight symptom}, which occurs when the weights in the network the output are not changing for several iterations. In that case, DD would identify the root cause as that the \textit{learning rate is too low} or that \textit{the optimizer is incorrect} and recommend a fix. 


In summary, this paper makes the following contributions:

\begin{itemize}
    \item We study different types of symptoms and propose a dynamic analysis  for detecting errors and recommending fixes.
    \item We introduced DeepDiagnosis (DD) the reference implementation of our approach. 
    \item We evaluated DD against SoTA. We found that DD is more efficient than UMLAUT~\cite{schoop2021umlaut} and DeepLocalize~\cite{wardat21DeepLocalize}. Also, DD has better performance than AUTOTRAINER~\cite{ZhangAUTOTRAINER}.
    \item We provide a set of 444 models that practitioners can use to evaluate their fault localization approaches.
    \item We make DD available, its source code, evaluation results, and the problem solutions for 444 buggy models at~\cite{DDRepo}. 
\end{itemize}


	


The rest of the paper is organized as follows:
\secref{sec:background} describes the motivation of our approach. 
\secref{sec:Approach} describes our dynamic failure symptoms detection algorithm.
\secref{sec:EVALUATION} describes the evaluation of our approach compared with prior works.
\secref{sec:THREATSTOVALIDITY} discusses the threats to validity.
\secref{sec:relatedwork} discusses related works, and \secref{sec:conclusion} concludes 
and discusses future work.

\begin{figure*}[ht!]
	\centering
	\includegraphics[width=0.9\linewidth,trim={0cm 0.cm 0cm 0.0cm}]{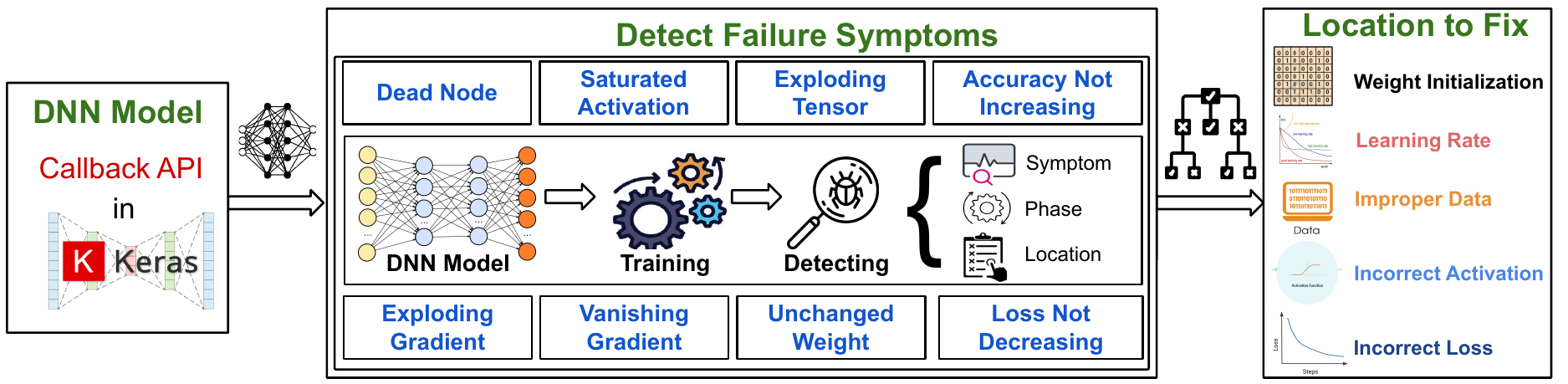}
    \caption{Overview of DeepDiagnosis.}
	\label{Process}
\end{figure*}

\section{A Motivating Example}
\label{sec:background}
\lstset{language=python}




In this section, we motivate our work by providing an example to illustrate the complexity of localizing faults and reporting their symptoms in DNN programs. 

\begin{lstlisting}[ language = python,label=motivation, caption=Bad Result for Simple Model~\cite{bug10}]
model = Sequential()
model.add(Dense(128, 50))
model.add(Activation('relu'))
model.add(Dropout(0.2))
model.add(Dense(50, 50))
model.add(Activation('relu'))
model.add(Dropout(0.2))
model.add(Dense(50, 1))
model.add(Activation('softmax'))
model.compile(loss='binary_crossentropy', optimizer=RMSprop())
model.fit(X,Y,batch_size,epoch,validation_data=(X_test,Y_test))
\end{lstlisting}

Consider the code snippet in~\lstref{motivation} from~\sof~\cite{bug10} with an example of a DNN. This model showed erratic behavior during training and returns bad results. At line 1, the developer constructed a sequential model and added a dense input layer at line 2 with the activation functions \texttt{relu} specified at line 3. Then the developer added a dropout layer at lines 4 and 7. Lines 5 and 8 are dense hidden layers with the activation functions \texttt{relu} and \texttt{softmax} specified at lines 6 and 9, respectively. The developer then compiled the model at line 10 and trained it using the \texttt{fit()} function at line 11. When compiling, the developer must specify additional properties, such as loss function and optimizer. In this example, the developer used as loss \texttt{binary\_crossentropy} and optimizer \texttt{RMSprop()} at line 10. Finally, at line~11, the developer specifies the training data, \texttt{batch\_size}, epoch, and \texttt{validation\_data}.

The developer noticed that the DNN program was providing bad accuracy and could not diagnose the problem nor fix it (\sof post~\cite{bug10}) while following the Keras MNIST example~\cite{Keras}.

The main issue with the code in~\lstref{motivation} is that it handles a \textit{binary classification} problem, and therefore it should \textbf{not} use the activation function \texttt{softmax} in line 9. As the \texttt{softmax} works for \textit{multi-class classifications} problems. Instead, it should use \texttt{sigmoid}, as it is the best suited for binary classification and will provide the best accuracy for the task.


\begin{table}[h!t]

    \caption{Result from Motivating Example}
  \label{tab:Result-motivate}
  \scalebox{0.80}{
\begin{tabular}{|c|l|l}
\cline{1-2}
\textbf{Approach}      & \multicolumn{1}{c|}{\textbf{Ouput}}                                                                                                                                                                                                                                   &  \\ \cline{1-2}
\textbf{UMLAUT}        & No Output &  \\ \cline{1-2}
\textbf{DeepLocalize}  & layer 7: Numerical Error in delta Weights   &  \\ \cline{1-2}
\textbf{AUTOTRAINER}  & \begin{tabular}[c]{@{}l@{}} solution,times,issue_list,train_result,describe\\
1. selu,0,['relu'],0.5,Using 'SeLU' activation in each layers'  \\
2. bn,0,['relu'],0.5,Using 'BatchNormalization' \\
...\\
Unsolved.. For more details~\cite{DDRepo}\\
 \end{tabular}&  \\ \cline{1-2}
\textbf{DeepDiagnosis} & 
 \begin{tabular}[c]{@{}l@{}}Layer 7: Numerical Error in delta Weights\\ Change the activation function at layer: 8 \end{tabular} 
                                                                                                                                  &  \\ \cline{1-2}
\end{tabular}}
\end{table}




The current state-of-the-art for DNN fault localization is limited in terms of speed, accuracy, and efficiency. Table~\ref{tab:Result-motivate} summarizes the analysis results from three tools (DeepLocalize~\cite{wardat21DeepLocalize}, UMLUAT~\cite{schoop2021umlaut}, AUTOTRAINER~\cite{ZhangAUTOTRAINER}) and our approach DeepDiagnosis to diagnose the DNN model in~\lstref{motivation}. To apply UMLUAT for the above example, we made semantic changes that were validated by the authors~\cite{schoop2021umlaut}. After 104.65 seconds, the training was terminated, with UMLAUT not reporting any problems. To apply DeepLocalize, we followed the instructions in the \gh repository~\cite{myRepo}. DeepLocalize prints the following message after 2.14 seconds: ``Layer 7: Numerical Error in delta Weights.'' This message indicates that there is a numerical error in the backpropagation stage during training. Indicating fault location, but it does not help developers to fix the problem.
To apply AUTOTRAINER, we followed the instructions in the \gh repository~\cite{UMLAUTRepo}. After performing the training phase, AUTOTRAINER did not solve the problem and took 495.83 seconds. Specifically, AUTOTRAINER detects a Dying ReLU symptom, but it does not provide the fault location -- whether if it is in line 3 or 6. AUTOTRAINER tries to automatically fix the issue by trying different strategies (i.e., substituting activation functions, adding batch normalization layer, and substituting initializer), which, unfortunately, are unsuccessful. 

Our approach DeepDiagnosis correctly reports the fault location and its symptoms after 35.03 seconds. Also, it provides a suggestion to perform a fix in the form of a message. Specifically, DeepDiagnosis reports that the bug is located in the backpropagation stage of layer 7 at line 8. Also, it prints out a numerical message: ``Error in delta Weights'', which indicates the type of the symptom. It also reports that the root cause is the activation function in layer 8 at line 9 (\texttt{softmax}). Finally, it answers the developer's question -- there is indeed a problem with the activation function in the last layer and not in the training dataset.

\section{Approach}
\label{sec:Approach}
In this section, we provide an overview of our approach for fault localization. We provide descriptions of failure symptoms and their root causes. Also, we describe the process of mapping symptoms to their root causes.

Our approach monitors the key values during training, like weights and gradients. During training, it analyzes the recorded value to detect symptoms and determine whether a training problem exists. If a symptom is detected, our approach invokes a Decision Tree to diagnose/repair information based on a set of pre-determined rules. Otherwise, the training will terminate with the trained model and reported the model is correct.

\begin{table*}[htbp]
\footnotesize
     \setlength{\parskip}{.01cm}
	\setlength{\belowcaptionskip}{.05cm}
  \caption{Methods for Detecting Failure Symptoms}
  \scalebox{0.98}{
    \begin{tabular}{|l|c|p{5.5em}|c|p{45.em}|}
    \hline
\cline{1-2}\cline{5-5}    \multicolumn{1}{|p{1.335em}|}{\textbf{ID}} & \multicolumn{1}{c|}{\textbf{Method Name}} & \multicolumn{1}{c|}{\textbf{Input}} & \multicolumn{1}{c|}{\textbf{Output }} & \multicolumn{1}{c|}{\textbf{Description}} \\
    \hline
    \rowcolor[rgb]{ .859,  .859,  .859}    S1& ExplodingTensor () & Weight, 
$\Delta$ Weight, 
and Layer Output & T $\mid$ F & The procedure detects any numerical error such as infinite, NaN (Not a Number), or zero. To that end, it computes the input's mean value. Then, it checks for a numerical error is detected. In case of error, it returns \textbf{True}, otherwise \textbf{False}.
\\
   \hline
  S2&
UnchangeWeight () &
Weight, 
$\Delta$ Weight, 
Layer Output &
T $\mid$ F & The procedure stores the value for a given number of steps (N = 5). Then it compares the value for the current step with the mean value stored in for previous (N = 5) steps. The evaluation takes place for every given number of steps. The procedure returns \textbf{True} if the value is not changing, otherwise \textbf{False}.
\\
\hline

\rowcolor[rgb]{ .859,  .859,  .859} S3& SaturatedActivation () & Input of Activation Function & T $\mid$ F &
The procedure detects if the tanh or sigmoid activation functions, or other logistic functions are becoming saturated. It does so by checking if their input has reached either a maximum or minimum value. Saturated functions' derivatives would be equal to zero at those points. The procedure counts the activity of a close or greater node than to the (Max_Threshold = 5) or less than (Min_Threshold = -5) of the activation function; If the percentage of total activity nodes is greater than the (Threshold_Layer = 0.5) percent of the nodes are saturated, the procedure returns \textbf{True}, otherwise \textbf{False}.
 \\
 \hline
  
S4& DeadNode () & Relu Output & T $\mid$ F & This procedure takes the output of Rectified Linear Unit (ReLU) activation function as input, then computes how many inactive nodes dropped below (Threshold = 0.0). If the percentage of inactive nodes is greater than (Layer_Threshold = 0.7) it returns \textbf{True}, otherwise \textbf{False}. \\
    \hline

\rowcolor[rgb]{ .859,  .859,  .859} S5&
OutofRange () &
Output of last layer &
T $\mid$ F &
The procedure detects if the activation function's output is becoming out of range for the labeling training dataset Y. To that end, it finds the range (maximum and minimum) of the activation function's output. Then compare it with Y labeling data. If the value is out of the boundary, the procedure returns \textbf{True}, otherwise \textbf{False}.\\
    \hline
    
S6& LossNotDecreasing () & Loss Value & T $\mid$ F & The procedure stores the loss value for every number of steps (N = 5), then compares the loss value for the current step with the mean value of losses stored in the previous (N = 5) steps. The evaluation happens for every number of steps (N = 5). The procedure returns \textbf{True} if the loss is not decreasing, otherwise \textbf{False}.   \\
    \hline
\rowcolor[rgb]{ .859,  .859,  .859} S7& AccuracyNotIncreasing () & Accuracy Value & T $\mid$ F & The procedure stores the accuracy value for every number of steps (N = 5), then compares the accuracy value for the current step with the mean value of accuracy stored in previous (N = 5) steps. The evaluation happened every number of steps (N = 5). The procedure returns \textbf{True} if accuracy is not increasing, otherwise \textbf{False}. \\
    \hline
S8& VanishingGradient () & Delta Weight & T $\mid$ F & This procedure detects the Vanishing Gradient problem by checking the gradients when they become extremely small or drop to zero. The procedure computes the mean of the gradients' absolute values, then checks if their means drop below a specified (Threshold = 0.0000001). In the case of a positive detection, it returns \textbf{True}, otherwise \textbf{False}. \\
\hline

    \end{tabular}}%
  \label{tab:addlabel}%

  \begin{tablenotes}
	\small
	\item This table shows procedures descriptions from~\cite{SageMaker, LearningRate, schoop2020scram, goodfellow2016deep, ZhangAUTOTRAINER}. T$\mid$F  indicates that the procedure returns True$\mid$ False respectively. \\[1ex]
	
	\end{tablenotes}
\end{table*}%

\begin{table*}[htbp]
\footnotesize
  \caption{Methods for Mapping from Failure Symptoms  to Location Fix}
  \scalebox{1.}{
    \begin{tabular}{|l|c|p{6.0em}|c|p{45.em}|}
    \hline
\cline{1-2}\cline{5-5}    \multicolumn{1}{|p{1.335em}|}{\textbf{No}} & \multicolumn{1}{c|}{\textbf{Method Name}} & \multicolumn{1}{c|}{\textbf{Input}} & \multicolumn{1}{c|}{\textbf{Output }} & \multicolumn{1}{c|}{\textbf{Description}} \\
    \hline
\rowcolor[rgb]{ .859,  .859,  .859} C1&
ImproperData () &
Training Data &
T $\mid$ F & Check if the maximum and minimum value of training dataset lie within specific range of [-1, 1].  If the value within the boundary, the procedure returns True. Otherwise, False.
\\
    \hline
C2& WeightInitialization () & Weight for each layer & T $\mid$ F &  This procedure checks the variance of weight inputs across layers to determine if a neural network has been poorly initialized. The procedure checks if the variance of weights per layer is equal or very close to 0 (Min_Threshold = 0.00001), or if it exceeds the (Min_Threshold = 10), the procedure returns \textbf{True}. Otherwise, \textbf{False}.\\
    \hline

\rowcolor[rgb]{ .859,  .859,  .859} C3& TuneLearn () & Learning rate, 
Weight, 
and $\Delta$ Weight & L $\mid$ H & The procedure evaluates the learning rate heuristically by computing the ratio of the norm of the gradient weight to the norm of weight for each layer. This ratio should be somewhere around (Learn_Threshold = 1e-3). If it is lower than (Learn_Threshold = 1e-3), then the learning rate might be too \textbf{Low}. If it is higher than (Learn_Threshold = 1e-3), the learning rate is likely too \textbf{High}. \\
\hline
    \end{tabular}}%
  \label{tab:checker}%
  
  \begin{tablenotes}
	\small
	\item This table is showing all the functionality of the procedures. T$\mid$F  indicates the procedure return True$\mid$ False respectively. L$\mid$H  indicates the procedure return Low$\mid$ High respectively. We borrowed these methods from existing literature~\cite{SageMaker, LearningRate, schoop2020scram, goodfellow2016deep, ZhangAUTOTRAINER} \\[1ex]
	
	\end{tablenotes}
\end{table*}%

\subsection{An Overview}


\fignref{Process} shows an overview of our approach for fault localization, DeepDiagnosis, and for suggesting locations fix. DD starts by receiving as input the initial model architecture with a training dataset and passing our callback method as a parameter to the \texttt{fit()} method (\fignref{Process} left component). This callback allows capturing and recording the key values (i.e., weight, gradient, etc.) during feed-forward and backward propagation stages (\fignref{Process} middle component). Then DD applies a dynamic detector during training to report different symptoms at different stages based on error conditions (see Section~\ref{sec:NumericalErrorDetection} for more details). If DD detects a symptom, it further analyzes the recorded key values to determine the input model's probable location for the fix (\fignref{Process} right component). Finally, DD reports the symptom type, which layers and stage the symptom was detected, and suggests a location fix.

\subsection{Failure symptoms and root causes}
\label{sec:NumericalErrorDetection}

Our goal is to detect failure symptoms as soon as possible during development. So that if the model is incorrect, developers would not have to wait until the end of the training to find that model has low accuracy, thus wasting computational resources. To that end, we collected 8 types of failure symptoms and their root causes from previous work in the AI research community~\cite{sussillo2014random, miller2018stable, isac2020effect, gulcehre2016noisy}. We provide more details of each of the symptoms and their root causes below.


\subsubsection{\textbf{Symptom \#1 Dead Node}} The Dead Node symptom takes place when most of a neural network is inactive. For example, assume that most of the neurons of a DNN are using the ReLU activation function, which returns zero when receiving any negative input. If the majority of the neurons receive negative values (e.g., due to a high learning rate), they would become inactive and incapable of discriminating the input. The DNN would end up with poor performance~\cite{ZhangAUTOTRAINER}. To identify this symptom, we compute the percentage of inactive neurons per layer. If the majority of the neural network is inactive, then we call it Dead Node.




\textbf{Root Causes:} This problem is likely to occur when~\cite{evci2018detecting}: (1) learning rate is too high/low. (2) there is a large negative bias. (3) improper weight or bias initialization. 

\subsubsection{\textbf{Symptom \#2 Saturated Activation}} The Saturated Activation symptom takes place when the input to the logistic activation function (e.g., \textit{tanh} or \textit{sigmoid}) reached either a very large or a very small value~\cite{gulcehre2016noisy}. At the saturated point, the function results would equal zero or be close to zero and, thus, leading to no weight updates. 
Our experiments show~\cite{DDRepo} that the behavior of \textit{sigmoid} and \textit{tanh} have a minimum saturated point at x=-5 and a maximum saturated point at x=5. Previous work showed that the saturated function affects the network's performance and makes the network difficult to train~\cite{gulcehre2016noisy,xu2016revise}. 



\textbf{Root Causes:} This problem is likely to occur when~\cite{glorot2010understanding}: (1) the input data are too large or too small; (2) improper weight or bias initialization; (3) learning rate is too high or too small. 

\subsubsection{\textbf{Symptom \#3 Exploding Tensor:}} The Exploding Tensor symptom takes place when the tensors' values become too large, leading to numerical errors in a feed-forward stage. For example, if the weight or output layer grows exponentially more than expected, becoming either infinite or \textit{NaN} (not a number). Eventually, this problem causes a numerical error, making it impossible for the model to learn.

\textbf{Root Causes:} This problem is likely to occur when~\cite{knox2018machine, goodfellow2016deep}:(1) the learning rate is too large; (2) there exist improper weight or bias initialization, or improper input data. 



\subsubsection{\textbf{Symptom \#4 Accuracy Not Increasing} \& \textbf{Symptom \#5 Loss Not Decreasing}} Both symptoms Accuracy Not Increasing and Loss Not Decreasing are very similar. The Accuracy Not Increasing symptom takes place when the accuracy of a target model is not increasing for N steps, but instead, it is decreasing or fluctuating during training. While for the Loss Not Decreasing symptom, the loss metric is the one that is not decreasing for N steps but is fluctuating. These behaviors indicate that the network will not achieve high performance. These symptoms are often caused by the incorrect selection of DNN hyperparameters~\cite{schoop2020scram}, such as loss function, activation function for the last layer, learning rate, optimizer, or batch size.



\textbf{Root Causes:} This problem is likely to occur when~\cite{knox2018machine, goodfellow2016deep}: (1) there exist improper training data; (2) the number of layers is too large/small;  and (3) the learning rate is very high/low; and (4) there exist incorrect activation functions.



\subsubsection{\textbf{Symptom \#6 Unchanged Weight}} The Unchanged Weight symptom takes place when the DNN weights do not have a noticeable influence on the output layers. This behavior leads to unchanging parameters and network stacks, which further prevents the model from learning~\cite{wardat21DeepLocalize, glorot2010understanding}. 



\textbf{Root Causes:} This problem is likely to occur when~\cite{wardat21DeepLocalize, glorot2010understanding}: (1) learning rate is very low; (2) the optimizer is incorrect; (3) there exist incorrect weights initialization; and (4) there exists incorrect loss/activation at the last layer. 

\subsubsection{\textbf{Symptom \#7 Exploding Gradient}} This problem occurs during the back-propagation stage. In it, gradients are growing exponentially from the last layer to the input layer, which leads to non-finite values, either infinite or \textit{NaN} (not a number). This issue makes learning unstable and sometimes even impossible. Consequently, updating the weights becomes very hard, and the training model ends up with a high loss or very low accuracy values.




\textbf{Root Causes:} This problem is likely to occur when~\cite{knox2018machine, goodfellow2016deep}:
(1) the learning rate is very high; (2) there is an improper weight or bias initialization; (3) there are improper  data input; and (4) the batch size is very large.

\subsubsection{\textbf{Symptom \#8 Vanishing Gradient}} The Vanishing Gradient problem occurs during the backward stage. When computing the gradient of the loss concerning weights using partial derivatives, the value of the gradient turns out to be so small or drops to zero. The problem causes major difficulty if it reaches the input layer, which will prevent the weight from changing its value during training. Since the gradients control how much the network learns during training, the neural network will end up without contributing to the prediction task or leading to poor performance~\cite{ZhangAUTOTRAINER,tan2019vanishing}.

\textbf{Root Causes:} This problem is likely to occur when~\cite{kong2017hexpo}: (1) the network has too many layers; (2) the learning rate is low; (3) the hidden layers improperly used \textit{Tanh} or \textit{Sigmoid}; and (4) there exists the incorrect weight initialization problem.


\subsection{Detecting Failure Symptoms}
\label{sec:Algorithem}
In Table~\ref{tab:addlabel} from Method S1 to S8, we describe the failure symptoms discussed in Section~\ref{sec:NumericalErrorDetection}, using its name, input/output, and the detection procedure. Algorithm~\ref{alg:fault-localize} shows an example of a dynamic analysis procedure, which DeepDiagnosis uses to detect failure symptoms during training (Table~\ref{tab:addlabel} {\it Description}). Also, the Algorithm~\ref{alg:fault-localize} reports failure locations, such as in which layer and phases (i.e., feed-forward and backward propagation). In case a failure is detected, the algorithm will trigger the {\it Mapping()} procedure to identify the location in the original DNN source code. By doing so, it will localize the bug and determine the optimal fix.

At line 1, Algorithm~\ref{alg:fault-localize} iterates over the training epochs, with the training dataset divided into batches. Line 3 shows the division of the training dataset into a mini-batch. On lines 2-28, the algorithm runs one batch of the training dataset before updating the internal model parameters. The neural network can be divided into two stages: First, the forward stage, in which the algorithm performs dynamic analysis and symptom detection, including Numerical Error, Dead node, Saturated Activation, and Out of Range, at lines 4-12. Second, the backward stage, in which the algorithm performs dynamic analysis to detect additional symptoms, such as Numerical Error, Vanishing Gradient, and Unchanged weight at lines 23-28.


\subsubsection{Feed-forward stage} 

At lines 5 \& 6 of the Algorithm~\ref{alg:fault-localize}, it computes the output of a feed-forward before and after applying the activation function. At line 7, it invokes the \textit{ExplodingTensor()} procedure (S1 in Table~\ref{tab:addlabel}) to determine if the output contains a numerical error obtained from the output value before/after the applying activation function, respectively. If there is an error, the algorithm reports the NS message as shown in Table~\ref{tab:message}. Next, it invokes the Mapping() procedure from the decision tree in Figure~\ref{DecisionTable} by providing the symptom (NS), location,  stage (FW), and layer (L). The decision tree returns the best actionable fix for the model (see Section
~\ref{sec:CorrectLocation} for more details).


At line 8, the Algorithm~\ref{alg:fault-localize} invokes the \textit{UnchangeWeight()} (S2 in Table~\ref{tab:addlabel}) procedure to detect whether the output before/ after applying the activation function is no longer changing across steps. If the procedure indicates that the value does not change for N iterations, we follow~\cite{wardat21DeepLocalize} and set N=5. The \textit{UnchangeWeight()} procedure can be applied either to the output before/after the activation function. The algorithm reports the message UCS, as shown in Table~\ref{tab:message}. At line 9, the Algorithm invokes the SaturatedActivation () procedure (S3 in Table~\ref{tab:addlabel}) for the layer that has a logistic activation function (i.e., {\tt tanh} or {\tt sigmoid}) to determine if the layer is becoming saturated. This procedure takes two arguments, the value before applying the activation function (V\_1) and the name of the activation function (V\_2.name). If the procedure determines that the layer is saturated, the algorithm reports the message SAS as shown in Table~\ref{tab:message}. 


At line 10, the Algorithm~\ref{alg:fault-localize} invokes the \textit{DeadNode()} procedure (S4 in Table~\ref{tab:addlabel}) to check the layers that use the Rectified Linear Unit (ReLU) activation function. The goal is to determine if the output after applying the activation function has dropped below a threshold~\cite{ZhangAUTOTRAINER}. This procedure is invoked only after applying the activation function. The algorithm reports the message DNS as shown in Table~\ref{tab:message} when the error is detected. Similarly, at line 11, it invokes the \textit{OutofRange()} procedure (S5 in Table~\ref{tab:addlabel}) in the last layer. The goal is to determine if the developer has chosen the correct activation function. The algorithm reports the message ORS as shown in Table~\ref{tab:message} if the error is detected.



In lines 13 \& 15 the algorithm interprets and validates how well the model is doing by computing the loss and accuracy metrics, respectively. Then it determines if there is any numerical error in those metrics at lines 14 \& 16, respectively. The algorithm invokes \textit{LossNotDecreasing()} and \textit{AccuracyNotIncreasing()} (S6 \& S7 in Table~\ref{tab:addlabel}) to check if the loss or the accuracy has not changed for a long time. In both cases, the algorithm reports a message LNDS or ANIS as shown in Table~\ref{tab:message}.



\subsubsection{Back propagation stage} 
During this stage, the Algorithm~\ref{alg:fault-localize} computes the gradient of loss function $\Delta$ Weight for the weight by chain rules in each iteration. At line 24, the algorithm invokes \textit{Backward()} to apply stochastic gradient descent, and this function returns the Weight and $\Delta$ Weight in each iteration. At line 25, the algorithm invokes the \textit{VanishingGradient()} procedure (S8 in Table~\ref{tab:addlabel}) and passes $\Delta$ Weight to check if the gradients become extremely small or close to being zero. In the same way, at line 26, the algorithm can determine if there is a numerical error in the Weight or the gradient weight in each layer by invoking the \textit{ExplodingTensor()} procedure (S1 in Table~\ref{tab:addlabel}). The backpropagation algorithm works if the Weight is updated using the gradient method and the loss value keeps reducing, to check if the backpropagation works effectively. In the backward propagation, we also invoke the \textit{UnchangeWeight()} procedure (S2 in Table~\ref{tab:addlabel}) to detect whether the weight or $\Delta$ Weight is no longer changing across steps. If any procedure decides that there is an issue, then the algorithm will return a message to indicate the type of symptom as shown in Table~\ref{tab:message}, L represents a faulty layer number. Then the algorithm invokes \textit{Mapping()} and passes the symptom, location, and layer to find the best actionable change to fix the issue in the model. Finally, if the algorithm did not detect any type of symptom, it will terminate after finishing the training at line 29 and print a message indicating that there is no issue in the model (CM).

\begin{table}
	\caption{Abbreviation of Actionable Changes}
	\label{tab:freq}
    \setlength{\parskip}{.2cm}
    \setlength{\intextsep}{1cm plus .1cm minus 1.cm} 
	\setlength{\belowcaptionskip}{.3cm}
	\centering
	\begin{tabular}{c|l|l}
		\hline\hline
		No & Message Guideline & Abbreviation\\
		\hline
		1 & Improper Data & MSG0\\
		\rowcolor[rgb]{ .859,  .859,  .859} 
		2 & Change the loss function  & MSG1\\
		3 & Change the activation function & MSG2\\
		\rowcolor[rgb]{ .859,  .859,  .859}
		4 & Change the learning rate & MSG3\\
		5 & Change the initialization  of weight  & MSG4\\
		\rowcolor[rgb]{ .859,  .859,  .859}
		6 & Change the layer number & MSG5\\
		7 & Change the optimizer & MSG6\\
		\hline
	\end{tabular}
\end{table}

\begin{table}
	\caption{Abbreviation of Failure Symptoms}
	\label{tab:message}
    \setlength{\parskip}{.2cm}
	\setlength{\intextsep}{1cm plus .1cm minus 1.cm} 
	\setlength{\belowcaptionskip}{.1cm}
	\centering
	\begin{tabular}{c|l|l}
		\hline\hline
		No & Symptoms & Abbreviation\\[0.5ex]
		\hline
		1 & Numerical Errors & NS\\
		\rowcolor[rgb]{ .859,  .859,  .859} 2 & Unchanged weight & UCS\\
		3 & Saturated Activation  & SAS\\
		\rowcolor[rgb]{ .859,  .859,  .859} 4 & Dead Node  & DNS\\
		5 & Out of Range   & ORS\\
		\rowcolor[rgb]{ .859,  .859,  .859} 6 & Loss Not Decreasing  & LNDS\\
		7 & Accuracy Not Increasing & ANIS\\
		\rowcolor[rgb]{ .859,  .859,  .859}  8 & Vanishing Gradient  & VGS\\
		9 & Invalid Loss  & ILS\\
		\rowcolor[rgb]{ .859,  .859,  .859} 10 & Invalid Accuracy & IAS\\
		11 & Correct Model  & CM\\
		\hline
	\end{tabular}
\end{table}
\begin{algorithm}
	\caption{Failure Symptoms Detection}
	\label{alg:fault-localize}
	\small
	\DontPrintSemicolon
	\SetKwData{Left}{left}
	\SetKwData{Up}{up}
	\SetKwFunction{Forward}{Activation}
	\SetKwInOut{Input}{input}
	\SetKwInOut{Output}{output}
	
	\Indm\Indmm
	\Input{Training data (input, label),  DNN program}
	\Output{Failure symptoms and locations (layers, iterations, epoch)}
	\Indp\Indpp
	\BlankLine
	\For{$e\leftarrow 0$ \KwTo $epochs$}{ 
		\For{$i\leftarrow 0$ \KwTo $Length(input)$ \KwStep $batchsize$ }{
			$X \leftarrow input[i]$; 
			$Y \leftarrow label[i]$\;
			\For{$L\leftarrow 0$ \KwTo $Length(Layers)$  }{
				$V_1$ $\leftarrow$ $Layer[L].Forward(X)$\;
				$V_2= Layer[L].Activation(V_1)$\;
				\lIf{$ExplodingTensor(V_2|V1)$}
				{
					\Return NS, $Mapping(NS, FW, L)$
				}
				\lIf{$UnchangeWeight(V_2|V_1)$}
				{
					\Return UCS, $Mapping(UCS,FW, L)$
				}
				\lIf{$Saturated(V_1,V_2.name)$}
				{
					\Return SAS, $Mapping(SAS,FW,L)$
				}
				
				\lIf{$DeadNode(V_2)$}
				{
					\Return DNS, $Mapping(DNS, FW, L)$
				}
			
				\lIf{$OutofRange(V_2, Y)$ \&\& $L == Last$}
				{
					\Return ORS, $Mapping(ORS, FW, L)$
				}
				
				$X \leftarrow V_2$\label{algo2:7}
				
			}
			$Loss  \leftarrow ComputeLoss(V_2, Y)$\;
			\lIf{$Loss$ is equal to NaN OR $inf$}
			{
				\Return ILS, $Mapping(ILS)$ \label{algo2:44}
			}
			$Accuracy \leftarrow ComputeAccuracy(V_2, Y)$\;
			\If{$Accuracy$ is equal to NaN OR inf OR $0$  }
			{
				\Return IAS, $Mapping(IAS)$ \label{algo2:44}
			}
			
			\If{$LossNotDecreasing(Loss)$}
			{
				\Return LNDS, $Mapping(LNDS)$ \label{algo2:44}
			}
			
			\If{$AccuracyNotIncreasing(Accuracy)$}
			{
				\Return ANIS, $MappingSymptoms(ANIS)$ \label{algo2:44}
			}
			$dy \leftarrow Y$\;
			\For{ $L \leftarrow Length(Layers)$ \KwTo $0$} {
				$V_3,  W[L] \leftarrow Layer[L].Backward(dy)$\; 
				\lIf{$VanishingGradient( W[L])$}
				{
					\Return VGS, $Mapping(VGS, BW, L)$
				}
				\lIf{$ExplodingTensor(V_3|W[L])$}
				{
					\Return NS, $Mapping(NS, V_3|DW, L)$
				}
				\lIf{$UnchangeWeight(V_3|W[L])$}
				{
					\Return UCS, $Mapping(UCS, V_3|DW, L)$
				}
				$dy \leftarrow V_3$\;
			}
		}
	}
	\Return CM\;
	\BlankLine
	\Indm\Indmm
	\Indp\Indpp
\end{algorithm}
\DecMargin{1.0em}

			
			
		
			    


\subsection{Mapping Symptoms to Location fix}
\label{sec:CorrectLocation}

\begin{figure*}[ht]
	\centering
	\includegraphics[width=0.99\linewidth,trim={0cm 0.cm 0cm 0.0cm},clip]{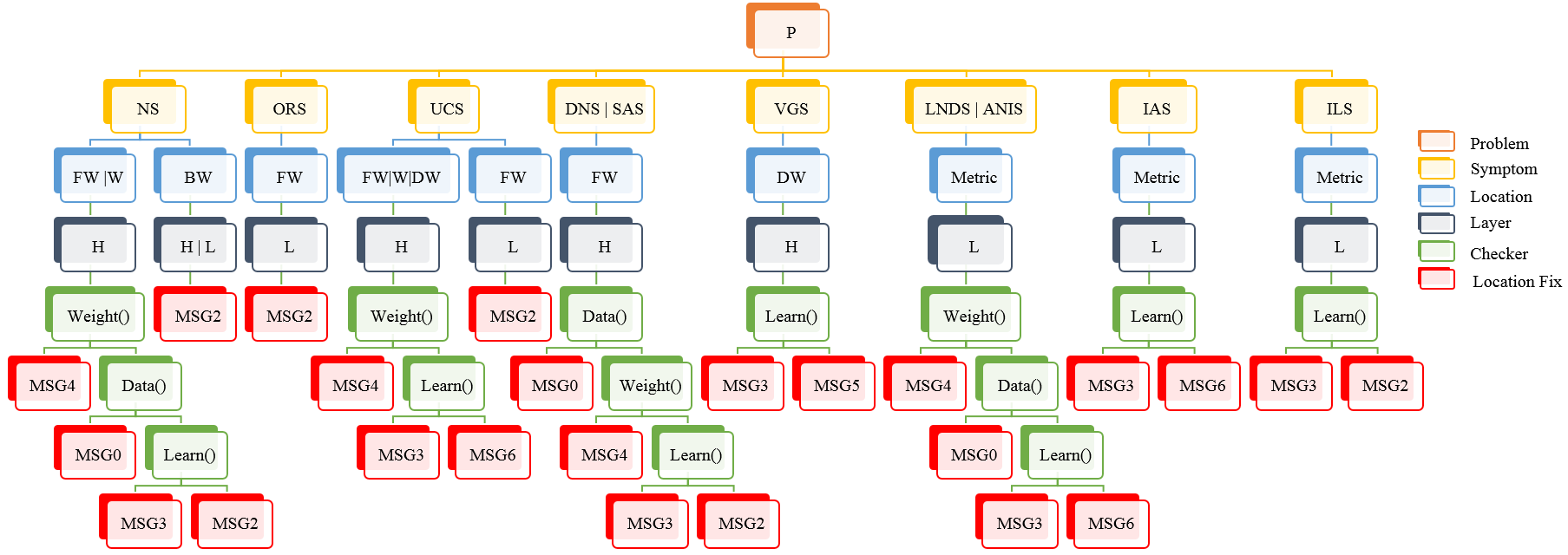}
	\caption{Mapping Symptoms to Fix Location}
	\label{DecisionTable}
\end{figure*}

\paragraph{Decision Tree:} The main goal of this step is to mitigate manual effort and reduce the time for diagnosing and fixing bugs. To that end, the \textit{Mapping()} procedure in Algorithm 1 provides fix suggestions based on the detected failure symptoms. Figure~\ref{DecisionTable} shows a representation of the Decision Tree which the \textit{Mapping()} procedure uses to provide a fix recommendation. 

The Decision Tree consists of 24 rules, which corresponds to decision paths. Each rule provides a mapping from failure symptoms and detected locations to actionable changes. The tree defines a binary classification rule which maps instances in the format problem (Symptom, Location, Layer) into one of seven classes of changes (Table~\ref{tab:freq}). The root node represents the problem, orange nodes the symptoms, blue nodes the locations, gray nodes the layer type, green nodes, the conditions, and red nodes the actionable changes. Table~\ref{tab:checker} shows the methods {\tt Data()}, {\tt Weight()} and {\tt Learn()}, which are used to compute the conditions. Each Decision Tree instance maps a path from the root to one of the leaves.

For example, assume that a developer wants to check the code in Listing~\ref{motivation}. To that end, the developers can use the Algorithm~\ref{alg:fault-localize} to verify the model. The algorithm invokes the \textit{Mapping()} procedure (line 26) by passing the symptom NS, location, stage BW (backward), and layer (7). This procedure traverses the path under the NS node in the Decision Tree (Figure~\ref{DecisionTable}). Since the problem occurred in the BW stage,  the algorithm takes the right path to satisfy the condition. Then, it verifies the layer type (7). Since it finds an issue in the layer, the procedure returns the message MSG2 -- Change the activation function (Table~\ref{tab:freq}).




\paragraph{Heuristics:} We developed a set of heuristics based on the root causes (see Section~\ref{sec:NumericalErrorDetection}). There are three main root causes: (1) Data Preparation; (2) Parameter Tuning; and (3) Model Architecture. For Data Preparation, the algorithm checks if the data is normalized (C1 - \textit{ImproperData()} in Table~\ref{tab:checker}). 
For Parameter Tuning, our approach checks if the hyperparameters (such as learning rate) were assigned correctly. Also, to check if the weights were initialized correctly, the algorithm invokes the \textit{WeightInitialization()}. The \textit{TuneLearn()} procedure verifies whether the learning rate is very high or very low (C2, and C3 in Table~\ref{tab:checker}, respectively). For model architecture, the algorithm searches for a relation between the location and the stage of the symptom. It performs this step to pinpoint which APIs are being misused in the model (e.g., loss, activation function). 

We collected the root causes for each symptom from previous work~\cite{sussillo2014random, miller2018stable, isac2020effect, gulcehre2016noisy} (more details in Section~\ref{sec:NumericalErrorDetection}). To arrive at a possible fix for a given symptom, we choose the most frequent root cause. We follow this approach as our findings show that changes in the order we check for the possible root causes do not affect the results, only on the total time to arrive at a solution. For example, assume that a model has the Dead Node symptom. 
In terms of frequency, improper data tends to happen more often than weight and learning rate. If the three root causes are correct, our approach checks the model architecture, which is the least common in this case. Thus arriving at an improper activation function as the root cause of this symptom. 

 {
 \color{blue}
 }

\section{Evaluation}
\label{sec:EVALUATION}



In the evaluation, we answer the following research questions:


\begin{itemize}
	\item RQ1 (Validation): Can our technique localize the faults and report the symptoms in Deep Learning programs effectively?
	\item RQ2 (Comparison): How does our technique for fault localization compared to existing methodologies in terms of time and effectiveness?
	\item RQ3 (Limitation): In which cases does our technique fail to localize the faults and report the symptoms? 
	\item RQ4 (Ablation): To what extent does each type of symptom we developed contribute to the overall performance of DeepDiagnosis? 
\end{itemize}

\subsection{Experimental setup}

\subsubsection{Implementation}

We implemented DeepDiagnosis on top of \keras 2.2.0~\cite{Keras} and \textit{TensorFlow} 2.1.0~\cite{TensorFlow}. Also, we implemented Algorithm~\ref{alg:fault-localize} by overriding the method called \textit{(on\_epoch\_end(epoch, logs=None)}. For the Decision Tree in Figure~\ref{DecisionTable}, we implemented it as a decision rule consisting of a set of conditional statements. The overridden method invokes the decision tree once upon detecting a symptom. Then it passes the symptom type as a parameter for the decision tree.


We conducted all the experiments on a computer with a 4 GHz Quad-Core Intel Core i7 processor and 32 GB 1867 MHz DDR3 GB of RAM running the 64-bit iMac version 4.11.

\subsubsection{Benchmark} In total, we collected 548 models from prior work~\cite{ZhangAUTOTRAINER,myRepo,schoop2021umlaut}. From these, we removed 104 RNN models, as our approach does not support them. The resulting 444 models are composed of 53, which are known to \textbf{have bugs} from~\cite{myRepo,schoop2021umlaut}. We refer to these 53 models SGS benchmark as they come from \textbf{S}tackOverflow, \textbf{G}itHub, and \textit{\textbf{S}choop}~\etal~\cite{schoop2021umlaut}. Also, the 391 models from~\cite{ZhangAUTOTRAINER} are in the compiled *.h5 format. The remaining 391 models are divided into two sets. In particular, the first with 188 correct models -- \textbf{without} any known bugs -- and the second with 203 buggy models -- \textbf{with} bugs.

Most machine learning developers share the source code or the trained weights of their models in *.h5 format. To allow others to improve the understanding of how a model operates and inspect it with new data, we implemented the \textit{Extractor} tool~\cite{DDRepo}. It extracts the DNN source code from a *.h5 file. The \textit{Extractor} follows three steps to generate the Keras source code: first, it saves the model's layer information to a JSON file. Then, it generates the Abstract Syntax Tree (AST) from the JSON file. Finally, it converts the AST to the source code. 

To build the ground truth for the SGS benchmark, we manually reviewed all models and their respective answers, from \sof, and commits, from \gh. We perform this verification process to determine the exact bug location and its root causes. For the remaining 391 models - 203 \textbf{buggy} models and 188 \textbf{not buggy} models, we used our \textit{Extractor} to generate the source code for each model before/after performing a fix; we used the difflib~\cite{difflib} module to generate the diff file from the fixed model. We use the diff to locate the changes in the model, thus locating the root causes and the actual location of its corresponding fix.  We consider a model successfully repaired if its accuracy has improved after the fix. 



\subsubsection{Results Representation} Table~\ref{tab:result} shows the summarized evaluation results of the following approaches: UMLUAT~\cite{schoop2021umlaut}, DeepLocalize~\cite{wardat21DeepLocalize}, AUTOTRAINER~\cite{ZhangAUTOTRAINER}, and our approach DeepDiagnosis. Please refer to the reproducibility repository~\cite{DDRepo} for the complete table. The first column shows the source of the buggy model. The second column lists the model ID. The third column provided the \sof post \# and the model name from \gh repositories, collected by Wardat~\etal~\cite{wardat21DeepLocalize}, and also the name of the model introduced by Schoop~\etal~\cite{UMLAUTRepo} respectively. To compare our approach with the results generated from previous approaches, we reported the results in the following columns (from left to right): Time, Identify Bug (IB), Fault Localization (FL), Failure Symptom (FS), and Location Fix (LF), and Ablation (AB). Time, in seconds, reports how long each tool takes to report its results. The columns Identify Bug (IB) and Fault Localization (FL) show whether the approach successfully identifies the bug and the fault location. Failure Symptom (FS) and Location Fix (LF) columns show whether the tool correctly reports a symptom and an actionable change (model repair fix). Finally, the Ablation (AB) column shows which of the procedures listed in Table~\ref{tab:addlabel} detects the failure symptoms. Under each approach, the ``Yes'' and ``No'' status indicates whether it has successfully reported the target problem. Also, the ``---'' status denotes if the approach does not yet support the target problem. Lastly, we use method ID in Table~\ref{tab:addlabel} to indicate which procedure is used to detect the failure symptom. 


Table~\ref{tab:datasetResult} summarizes the analysis results from four approaches using benchmarks collected from three different sources~\cite{wardat21DeepLocalize, schoop2021umlaut, ZhangAUTOTRAINER}. The second column (Total) lists the total number of buggy models for each dataset. To compare our approach with previous approaches, we reported Time, in seconds, the average time  for each tool takes to report its results for each dataset. The columns Identify Bug (IB) shows how many each approach successfully identifies the bug from each dataset. Our approach is capable of handling eight types of symptoms with different
types of datasets using different types of model architectures. Table~\ref{tab:SymptomsDD} shows the number of symptoms detected from different types of datasets.

\begin{table*}[htbp]
  
  \centering
   \setlength{\parskip}{.01cm}
	\setlength{\belowcaptionskip}{.05cm}
	
  \caption{Comparing the Results from UMLAUT (UM), DeepLocalize (DL), AUTOTRAINER (AT) and  DeepDiagnosis (DD)}
  \scalebox{0.70}{
	
    \begin{tabular}{|c|r|l|r|r|r|r|r|r|c|c|c|c|c|c|c|c|c|c|c|c|c|c|c|} 
    \hline
    \multirow{2}[4]{*}{} & \multicolumn{1}{c|}{\multirow{2}[4]{*}{\textbf{No}}} & \multicolumn{1}{c|}{\multirow{2}[4]{*}{\textbf{Post \#}}} & \multicolumn{4}{c|}{\textbf{Time}} &  \multicolumn{4}{c|}{\textbf{Identify Bug (IB)}} & \multicolumn{4}{c|}{\textbf{Fault Localization (FL)}} & \multicolumn{4}{c|}{\textbf{Failure Symptom (FS)}} & \multicolumn{4}{c|}{\textbf{Location Fix (LF)}} & \multicolumn{1}{c|}{{\textbf{AB}}}\\
\cline{4-23}          &       &       & 
\multicolumn{1}{c|}{\textbf{\textbf{UM }}} & \multicolumn{1}{c|}{\textbf{\textbf{DL}}} & \multicolumn{1}{c|}{\textbf{AT}}& \multicolumn{1}{c|}{\textbf{DD}} &
\multicolumn{1}{c|}{\textbf{UM }} & \multicolumn{1}{c|}{\textbf{DL}}& \multicolumn{1}{c|}{\textbf{AT}} & \multicolumn{1}{c|}{\textbf{DD}} & 
\multicolumn{1}{c|}{\textbf{UM }} & \multicolumn{1}{c|}{\textbf{DL}} & \multicolumn{1}{c|}{\textbf{AT}}& \multicolumn{1}{c|}{\textbf{DD}} & 
\multicolumn{1}{c|}{\textbf{UM }} & \multicolumn{1}{c|}{\textbf{DL}} & \multicolumn{1}{c|}{\textbf{AT}}& \multicolumn{1}{c|}{\textbf{DD}} &
\multicolumn{1}{c|}{\textbf{UM }} & \multicolumn{1}{c|}{\textbf{DL}} & \multicolumn{1}{c|}{\textbf{AT}}& \multicolumn{1}{c|}{\textbf{DD}} &\\
    \hline
    \hline
    
    \multirow{4}[6]{*}{\begin{turn}{-7166}\textbf{\textbf{\sof \cite{myRepo}}}\end{turn}} 
                      &\cellcolor[rgb]{ .859,  .859,  .859} 1  & \multicolumn{1}{r|}{48385830\cellcolor[rgb]{ .859,  .859,  .859}} & \cellcolor[rgb]{ .859,  .859,  .859} 0.39	& \cellcolor[rgb]{ .859,  .859,  .859} 2.14	& \cellcolor[rgb]{ .859,  .859,  .859} 103.91 & \cellcolor[rgb]{ .859,  .859,  .859} 8.27	& \cellcolor[rgb]{ .859,  .859,  .859} Yes	& \cellcolor[rgb]{ .859,  .859,  .859} Yes	& \cellcolor[rgb]{ .859,  .859,  .859} Yes & \cellcolor[rgb]{ .859,  .859,  .859} Yes	& \cellcolor[rgb]{ .859,  .859,  .859} Yes	& \cellcolor[rgb]{ .859,  .859,  .859} Yes	& \cellcolor[rgb]{ .859,  .859,  .859} --- &
                      \cellcolor[rgb]{ .859,  .859,  .859} Yes	& \cellcolor[rgb]{ .859,  .859,  .859} Yes	& \cellcolor[rgb]{ .859,  .859,  .859} Yes	&
                      \cellcolor[rgb]{ .859,  .859,  .859} Yes & \cellcolor[rgb]{ .859,  .859,  .859} Yes	& \cellcolor[rgb]{ .859,  .859,  .859} Yes	& \cellcolor[rgb]{ .859,  .859,  .859} No	& \cellcolor[rgb]{ .859,  .859,  .859} No & \cellcolor[rgb]{ .859,  .859,  .859} Yes	
                      & \cellcolor[rgb]{ .859,  .859,  .859} \#1\\
                      
\cline{2-24}          & 2  & \multicolumn{1}{r|}{44164749} &188.61	&111.56	& 197.90 &242.34	&No	&Yes& No	&No	&No	&Yes& ---	&No	&No	&Yes & No	&No	&No	&No	& No &No  &	---\\ 

\cline{2-24}          & \cellcolor[rgb]{ .859,  .859,  .859} 3  &   \multicolumn{1}{r|}{\cellcolor[rgb]{ .859,  .859,  .859} 31556268} & \cellcolor[rgb]{ .859,  .859,  .859}  ---	& \cellcolor[rgb]{ .859,  .859,  .859} 1.2	& \cellcolor[rgb]{ .859,  .859,  .859} --- & \cellcolor[rgb]{ .859,  .859,  .859} 12.48	&  \cellcolor[rgb]{ .859,  .859,  .859}  ---	& \cellcolor[rgb]{ .859,  .859,  .859} Yes	& \cellcolor[rgb]{ .859,  .859,  .859} --- & \cellcolor[rgb]{ .859,  .859,  .859} Yes	& \cellcolor[rgb]{ .859,  .859,  .859}  ---	& \cellcolor[rgb]{ .859,  .859,  .859} No & \cellcolor[rgb]{ .859,  .859,  .859} --- 	& \cellcolor[rgb]{ .859,  .859,  .859} Yes	& \cellcolor[rgb]{ .859,  .859,  .859}  ---	& \cellcolor[rgb]{ .859,  .859,  .859} Yes & \cellcolor[rgb]{ .859,  .859,  .859} --- 	& \cellcolor[rgb]{ .859,  .859,  .859} Yes	& \cellcolor[rgb]{ .859,  .859,  .859}  ---	& \cellcolor[rgb]{ .859,  .859,  .859} No & \cellcolor[rgb]{ .859,  .859,  .859} --- 	& \cellcolor[rgb]{ .859,  .859,  .859} Yes	& \cellcolor[rgb]{ .859,  .859,  .859} \#7\\ 

\cline{2-24}          & 4  & \multicolumn{1}{r|}{50306988}  &1.9	&3.57& 93.60	&1.75	&No	&Yes & Yes	&Yes	&No	&Yes& ---	&Yes	&No	&Yes& Yes	&Yes	&No	&No	& Yes &Yes &	\#1\\ 

\cline{2-24}          & \cellcolor[rgb]{ .859,  .859,  .859} 5  & \multicolumn{1}{r|}{\cellcolor[rgb]{ .859,  .859,  .859} 48251943}  & \cellcolor[rgb]{ .859,  .859,  .859}  ---	& \cellcolor[rgb]{ .859,  .859,  .859} 706.83	&\cellcolor[rgb]{ .859,  .859,  .859} --- & \cellcolor[rgb]{ .859,  .859,  .859} 1.61	& \cellcolor[rgb]{ .859,  .859,  .859}  ---	& \cellcolor[rgb]{ .859,  .859,  .859} No & \cellcolor[rgb]{ .859,  .859,  .859} --- 	& \cellcolor[rgb]{ .859,  .859,  .859} Yes	& \cellcolor[rgb]{ .859,  .859,  .859}  ---	& \cellcolor[rgb]{ .859,  .859,  .859} No	&\cellcolor[rgb]{ .859,  .859,  .859} --- & \cellcolor[rgb]{ .859,  .859,  .859} Yes	& \cellcolor[rgb]{ .859,  .859,  .859}  ---	& \cellcolor[rgb]{ .859,  .859,  .859} No	&\cellcolor[rgb]{ .859,  .859,  .859} --- & \cellcolor[rgb]{ .859,  .859,  .859} Yes	& \cellcolor[rgb]{ .859,  .859,  .859}  ---	& \cellcolor[rgb]{ .859,  .859,  .859} No	&\cellcolor[rgb]{ .859,  .859,  .859} --- & \cellcolor[rgb]{ .859,  .859,  .859} Yes & \cellcolor[rgb]{ .859,  .859,  .859} \#5	\\ 

\cline{2-24}          & 6  & \multicolumn{1}{r|}{38648195}  &5.4	&25.92& 85.38	&15.12	&Yes	&Yes& No	&Yes	&No	&Yes& --	&Yes	&No	&Yes& No	&Yes	&No	&No& No	&Yes&	\#1\\ 

    \hline
    \hline
    \multirow{7}[6]{*}{\begin{turn}{-7166}\textbf{\textbf{\gh \cite{myRepo}}}\end{turn}} 
                      &\cellcolor[rgb]{ .859,  .859,  .859} 7 & \cellcolor[rgb]{ .859,  .859,  .859}GH \#1  &\cellcolor[rgb]{ .859,  .859,  .859}128.67	&\cellcolor[rgb]{ .859,  .859,  .859}11.80	&\cellcolor[rgb]{ .859,  .859,  .859}6524.21&\cellcolor[rgb]{ .859,  .859,  .859}	44.90	&\cellcolor[rgb]{ .859,  .859,  .859}Yes	&\cellcolor[rgb]{ .859,  .859,  .859}Yes	&\cellcolor[rgb]{ .859,  .859,  .859}Yes	&\cellcolor[rgb]{ .859,  .859,  .859}Yes	&\cellcolor[rgb]{ .859,  .859,  .859}No&\cellcolor[rgb]{ .859,  .859,  .859}	No&\cellcolor[rgb]{ .859,  .859,  .859}	 ---	&\cellcolor[rgb]{ .859,  .859,  .859}No&	\cellcolor[rgb]{ .859,  .859,  .859}No	&\cellcolor[rgb]{ .859,  .859,  .859}No&\cellcolor[rgb]{ .859,  .859,  .859}	Yes&\cellcolor[rgb]{ .859,  .859,  .859}	No	&\cellcolor[rgb]{ .859,  .859,  .859}No&	\cellcolor[rgb]{ .859,  .859,  .859}No&	\cellcolor[rgb]{ .859,  .859,  .859}Yes&	\cellcolor[rgb]{ .859,  .859,  .859}No
 &\cellcolor[rgb]{ .859,  .859,  .859} \#1	\\ 

\cline{2-24}          & 8 &GH \#2  & ---&	8432.06&	 ---&	1001.40&	 --- &	No	& ---&	No&	 ---&	No	& ---	&No&	 ---&	No&	 ---&	No&	 ---& 	No&	 ---&	No & \---	\\ 

\cline{2-24}          &\cellcolor[rgb]{ .859,  .859,  .859}9 & \cellcolor[rgb]{ .859,  .859,  .859}GH \#3  & \cellcolor[rgb]{ .859,  .859,  .859} ---&\cellcolor[rgb]{ .859,  .859,  .859}	31.69&\cellcolor[rgb]{ .859,  .859,  .859}	 ---&\cellcolor[rgb]{ .859,  .859,  .859}	2.17&\cellcolor[rgb]{ .859,  .859,  .859}	 ---&\cellcolor[rgb]{ .859,  .859,  .859}	Yes&\cellcolor[rgb]{ .859,  .859,  .859}	 ---&\cellcolor[rgb]{ .859,  .859,  .859}	Yes&\cellcolor[rgb]{ .859,  .859,  .859}	 ---&\cellcolor[rgb]{ .859,  .859,  .859}	Yes&\cellcolor[rgb]{ .859,  .859,  .859}	 ---	&\cellcolor[rgb]{ .859,  .859,  .859}Yes&\cellcolor[rgb]{ .859,  .859,  .859}	 ---&\cellcolor[rgb]{ .859,  .859,  .859}	No&	\cellcolor[rgb]{ .859,  .859,  .859} ---&\cellcolor[rgb]{ .859,  .859,  .859} 	Yes&\cellcolor[rgb]{ .859,  .859,  .859}	 ---&\cellcolor[rgb]{ .859,  .859,  .859} 	No	&\cellcolor[rgb]{ .859,  .859,  .859} ---& \cellcolor[rgb]{ .859,  .859,  .859}	Yes&	\cellcolor[rgb]{ .859,  .859,  .859} \#5\\ 

\cline{2-24}          &10 &GH \#4 &36.58&	102.44&	315.61&	102.96&	Yes&	Yes&	Yes&	Yes&	No&	No&	 ---&	No&	Yes&	No&	Yes&	No&	No&	No&	Yes&	No
&	\#4\\ 
 
\cline{2-24}          &\cellcolor[rgb]{ .859,  .859,  .859} 11 & \cellcolor[rgb]{ .859,  .859,  .859}GH \#5 &\cellcolor[rgb]{ .859,  .859,  .859} 18.95&\cellcolor[rgb]{ .859,  .859,  .859}	164.70&\cellcolor[rgb]{ .859,  .859,  .859}	173.92&	\cellcolor[rgb]{ .859,  .859,  .859}140.58&	\cellcolor[rgb]{ .859,  .859,  .859}Yes&\cellcolor[rgb]{ .859,  .859,  .859}	Yes&\cellcolor[rgb]{ .859,  .859,  .859}	No&\cellcolor[rgb]{ .859,  .859,  .859}	Yes&\cellcolor[rgb]{ .859,  .859,  .859}	No&\cellcolor[rgb]{ .859,  .859,  .859}	Yes&\cellcolor[rgb]{ .859,  .859,  .859}	 ---	&\cellcolor[rgb]{ .859,  .859,  .859}Yes&\cellcolor[rgb]{ .859,  .859,  .859}	No&\cellcolor[rgb]{ .859,  .859,  .859}	Yes&\cellcolor[rgb]{ .859,  .859,  .859}	No&\cellcolor[rgb]{ .859,  .859,  .859}	Yes&\cellcolor[rgb]{ .859,  .859,  .859}	No&\cellcolor[rgb]{ .859,  .859,  .859}	No&\cellcolor[rgb]{ .859,  .859,  .859}	No&\cellcolor[rgb]{ .859,  .859,  .859}	No &	\cellcolor[rgb]{ .859,  .859,  .859} \#2\\ 
 
\cline{2-24}          & 12 &GH \#6 &  ---&	9568.09&	12.57&	118.59&	 ---&	No&	No&	No&	 ---&	No&	 ---&	No&	 ---&	No&	no&	No&	 ---&	No&	No&	No& \---	\\

    \hline
    \hline
    \multirow{4}[6]{*}{\begin{turn}{-7166}\textbf{\textbf{Schoop~\etal~\cite{UMLAUTRepo}}}\end{turn}}
                      & 13 &A1 (C-10)&1.77&	18.39&	43.96&	2.75&	Yes&	Yes&	Yes&	Yes&	Yes&	Yes&	 ---&	Yes&	Yes&	Yes&	No&	Yes&	Yes&	No&	No&	Yes
&\#5	\\ 
   
\cline{2-24}          & \cellcolor[rgb]{ .859,  .859,  .859}14 & \cellcolor[rgb]{ .859,  .859,  .859}A2 (C-10)  &\cellcolor[rgb]{ .859,  .859,  .859} 1.50&	\cellcolor[rgb]{ .859,  .859,  .859}44.93&\cellcolor[rgb]{ .859,  .859,  .859}	18.36&	\cellcolor[rgb]{ .859,  .859,  .859}10.44&	\cellcolor[rgb]{ .859,  .859,  .859}Yes&	\cellcolor[rgb]{ .859,  .859,  .859}Yes&	\cellcolor[rgb]{ .859,  .859,  .859}No&\cellcolor[rgb]{ .859,  .859,  .859}	Yes&\cellcolor[rgb]{ .859,  .859,  .859}	No&	\cellcolor[rgb]{ .859,  .859,  .859}Yes&	 \cellcolor[rgb]{ .859,  .859,  .859}---&	\cellcolor[rgb]{ .859,  .859,  .859}Yes&\cellcolor[rgb]{ .859,  .859,  .859}	Yes&	\cellcolor[rgb]{ .859,  .859,  .859}Yes&	\cellcolor[rgb]{ .859,  .859,  .859}No&	\cellcolor[rgb]{ .859,  .859,  .859}No&\cellcolor[rgb]{ .859,  .859,  .859}	Yes&	\cellcolor[rgb]{ .859,  .859,  .859}No&	\cellcolor[rgb]{ .859,  .859,  .859}No&	\cellcolor[rgb]{ .859,  .859,  .859}No&\cellcolor[rgb]{ .859,  .859,  .859} \#1	\\ 

\cline{2-24}          & 15 &A3 (C-10)&348.88&	44.89&	119.54&	5.03&	Yes&	Yes&	Yes&	Yes&	No&	No&	 ---&	No&	Yes&	No&	No&	Yes&	Yes&	No&	No&	No
&	\#1\\ 
 
\cline{2-24}          & \cellcolor[rgb]{ .859,  .859,  .859}16 & \cellcolor[rgb]{ .859,  .859,  .859}B1 (C-10) &\cellcolor[rgb]{ .859,  .859,  .859}347.21&	\cellcolor[rgb]{ .859,  .859,  .859}10.65&	\cellcolor[rgb]{ .859,  .859,  .859}80.38&\cellcolor[rgb]{ .859,  .859,  .859}	2.17&	\cellcolor[rgb]{ .859,  .859,  .859}Yes&	\cellcolor[rgb]{ .859,  .859,  .859}Yes&	\cellcolor[rgb]{ .859,  .859,  .859}Yes&\cellcolor[rgb]{ .859,  .859,  .859}	Yes&	\cellcolor[rgb]{ .859,  .859,  .859}No&	\cellcolor[rgb]{ .859,  .859,  .859}No&	 \cellcolor[rgb]{ .859,  .859,  .859}---&	\cellcolor[rgb]{ .859,  .859,  .859}No&	\cellcolor[rgb]{ .859,  .859,  .859}Yes&\cellcolor[rgb]{ .859,  .859,  .859}	Yes&\cellcolor[rgb]{ .859,  .859,  .859}	Yes&	\cellcolor[rgb]{ .859,  .859,  .859}Yes&	\cellcolor[rgb]{ .859,  .859,  .859}Yes&\cellcolor[rgb]{ .859,  .859,  .859}	No&\cellcolor[rgb]{ .859,  .859,  .859}	No&	\cellcolor[rgb]{ .859,  .859,  .859}No
&	\cellcolor[rgb]{ .859,  .859,  .859} \#1\\ 
 
\cline{2-24}          & 17 &B2 (C-10) &3.42&	45.02&	16.90&	5.44&	Yes&	Yes&	No&	Yes&	No&	Yes&	 ---&	Yes&	Yes&	No&	No&	Yes&	Yes&	No&	No&	Yes&\#1	\\ 
 
\cline{2-24}          & \cellcolor[rgb]{ .859,  .859,  .859}18 & \cellcolor[rgb]{ .859,  .859,  .859}B3 (C-10) &1605.99\cellcolor[rgb]{ .859,  .859,  .859}&\cellcolor[rgb]{ .859,  .859,  .859}	45.54&\cellcolor[rgb]{ .859,  .859,  .859}	15.49&\cellcolor[rgb]{ .859,  .859,  .859}	15.49&\cellcolor[rgb]{ .859,  .859,  .859}	Yes&\cellcolor[rgb]{ .859,  .859,  .859}	Yes&	\cellcolor[rgb]{ .859,  .859,  .859}No&	\cellcolor[rgb]{ .859,  .859,  .859}No&\cellcolor[rgb]{ .859,  .859,  .859}	No&\cellcolor[rgb]{ .859,  .859,  .859}	No&	 \cellcolor[rgb]{ .859,  .859,  .859}---&\cellcolor[rgb]{ .859,  .859,  .859}	No&\cellcolor[rgb]{ .859,  .859,  .859}	Yes&	\cellcolor[rgb]{ .859,  .859,  .859}No&	\cellcolor[rgb]{ .859,  .859,  .859}No&\cellcolor[rgb]{ .859,  .859,  .859}	No&\cellcolor[rgb]{ .859,  .859,  .859}	Yes&	\cellcolor[rgb]{ .859,  .859,  .859}No&\cellcolor[rgb]{ .859,  .859,  .859}	No&\cellcolor[rgb]{ .859,  .859,  .859}	No
 &\cellcolor[rgb]{ .859,  .859,  .859} \ ---\\

    \hline

    \end{tabular}%
}
  \label{tab:result}%
   \begin{tablenotes}
	\small
	\item 
	C-10: indicates to the model using CIFAR-10 dataset, and F-M: indicates to the model using Fashion-MNIST dataset.
\end{tablenotes}
\end{table*}%

\begin{table}[]
\caption{Runtime Overhead vs. Problem Detects}
\scalebox{0.60}{
\begin{tabular}{|c|c|l|l|l|l|l|l|l|l|}
\hline
                                                            &                                                       & \multicolumn{4}{c|}{\textbf{Time}}                                                                                                                                                                                   & \multicolumn{4}{c|}{\textbf{Identify   Bug (IB)}}                                                                                                                                                   \\ \cline{3-10} 
\multirow{-2}{*}{\textbf{Dataset}}                          & \multirow{-2}{*}{\textbf{Total}}                      & \multicolumn{1}{c|}{\textbf{UM}}                 & \multicolumn{1}{c|}{\textbf{DL}}                 & \multicolumn{1}{c|}{\textbf{AT}}                   & \multicolumn{1}{c|}{\textbf{DD}}                          & \multicolumn{1}{c|}{\textbf{UM}}              & \multicolumn{1}{c|}{\textbf{DL}}             & \multicolumn{1}{c|}{\textbf{AT}}             & \multicolumn{1}{c|}{\textbf{DD}}                      \\ \hline
\cellcolor[HTML]{FFFFFF}                                    & \cellcolor[HTML]{FFFFFF}                              & \cellcolor[HTML]{FFFFFF}                         & \cellcolor[HTML]{FFFFFF}                         & \cellcolor[HTML]{FFFFFF}                           & \cellcolor[HTML]{FFFFFF}                                  & \cellcolor[HTML]{FFFFFF}                      & \cellcolor[HTML]{FFFFFF}                     & \cellcolor[HTML]{FFFFFF}                     & \cellcolor[HTML]{FFFFFF}                              \\
\multirow{-2}{*}{\cellcolor[HTML]{FFFFFF}\textbf{\sof~\cite{myRepo}}}      & \multirow{-2}{*}{\cellcolor[HTML]{FFFFFF}\textbf{29}} & \multirow{-2}{*}{\cellcolor[HTML]{FFFFFF}46.16}  & \multirow{-2}{*}{\cellcolor[HTML]{FFFFFF}421.39} & \multirow{-2}{*}{\cellcolor[HTML]{FFFFFF}771.56}   & \multirow{-2}{*}{\cellcolor[HTML]{FFFFFF}\textbf{103.74}}   & \multirow{-2}{*}{\cellcolor[HTML]{FFFFFF}10}  & \multirow{-2}{*}{\cellcolor[HTML]{FFFFFF}27} & \multirow{-2}{*}{\cellcolor[HTML]{FFFFFF}16} & \multirow{-2}{*}{\cellcolor[HTML]{FFFFFF}\textbf{26}} \\ \hline
\cellcolor[HTML]{C0C0C0}                                    & \cellcolor[HTML]{C0C0C0}                              & \cellcolor[HTML]{C0C0C0}                         & \cellcolor[HTML]{C0C0C0}                         & \cellcolor[HTML]{C0C0C0}                           & \cellcolor[HTML]{C0C0C0}                                  & \cellcolor[HTML]{C0C0C0}                      & \cellcolor[HTML]{C0C0C0}                     & \cellcolor[HTML]{C0C0C0}                     & \cellcolor[HTML]{C0C0C0}                              \\
\multirow{-2}{*}{\cellcolor[HTML]{C0C0C0}\textbf{\gh~\cite{myRepo}}}       & \multirow{-2}{*}{\cellcolor[HTML]{C0C0C0}\textbf{11}} & \multirow{-2}{*}{\cellcolor[HTML]{C0C0C0}46.16}  & \multirow{-2}{*}{\cellcolor[HTML]{C0C0C0}2613.6} & \multirow{-2}{*}{\cellcolor[HTML]{C0C0C0}148.41}   & \multirow{-2}{*}{\cellcolor[HTML]{C0C0C0}\textbf{137.82}} & \multirow{-2}{*}{\cellcolor[HTML]{C0C0C0}4}   & \multirow{-2}{*}{\cellcolor[HTML]{C0C0C0}7}  & \multirow{-2}{*}{\cellcolor[HTML]{C0C0C0}3}  & \multirow{-2}{*}{\cellcolor[HTML]{C0C0C0}\textbf{9}} \\ \hline
\cellcolor[HTML]{FFFFFF}                                    & \cellcolor[HTML]{FFFFFF}                              & \cellcolor[HTML]{FFFFFF}                         & \cellcolor[HTML]{FFFFFF}                         & \cellcolor[HTML]{FFFFFF}                           & \cellcolor[HTML]{FFFFFF}                                  & \cellcolor[HTML]{FFFFFF}                      & \cellcolor[HTML]{FFFFFF}                     & \cellcolor[HTML]{FFFFFF}                     & \cellcolor[HTML]{FFFFFF}                              \\
\multirow{-2}{*}{\cellcolor[HTML]{FFFFFF}\textbf{Schoop~\etal~\cite{UMLAUTRepo}}}   & \multirow{-2}{*}{\cellcolor[HTML]{FFFFFF}\textbf{12}} & \multirow{-2}{*}{\cellcolor[HTML]{FFFFFF}193.52} & \multirow{-2}{*}{\cellcolor[HTML]{FFFFFF}93.17}  & \multirow{-2}{*}{\cellcolor[HTML]{FFFFFF}3491.32}  & \multirow{-2}{*}{\cellcolor[HTML]{FFFFFF}\textbf{1020.20}}  & \multirow{-2}{*}{\cellcolor[HTML]{FFFFFF}12}  & \multirow{-2}{*}{\cellcolor[HTML]{FFFFFF}11} & \multirow{-2}{*}{\cellcolor[HTML]{FFFFFF}5}  & \multirow{-2}{*}{\cellcolor[HTML]{FFFFFF}\textbf{11}} \\ \hline
\cellcolor[HTML]{C0C0C0}                                    & \cellcolor[HTML]{C0C0C0}                              & \cellcolor[HTML]{C0C0C0}                         & \cellcolor[HTML]{C0C0C0}                         & \cellcolor[HTML]{C0C0C0}                           & \cellcolor[HTML]{C0C0C0}                                  & \cellcolor[HTML]{C0C0C0}                      & \cellcolor[HTML]{C0C0C0}                     & \cellcolor[HTML]{C0C0C0}                     & \cellcolor[HTML]{C0C0C0}                              \\
\multirow{-2}{*}{\cellcolor[HTML]{C0C0C0}\textbf{Blob~\cite{AUTORepo}}}     & \multirow{-2}{*}{\cellcolor[HTML]{C0C0C0}\textbf{48}} & \multirow{-2}{*}{\cellcolor[HTML]{C0C0C0}---}    & \multirow{-2}{*}{\cellcolor[HTML]{C0C0C0}113.14} & \multirow{-2}{*}{\cellcolor[HTML]{C0C0C0}112.6}    & \multirow{-2}{*}{\cellcolor[HTML]{C0C0C0}564.19}          & \multirow{-2}{*}{\cellcolor[HTML]{C0C0C0}---} & \multirow{-2}{*}{\cellcolor[HTML]{C0C0C0}44} & \multirow{-2}{*}{\cellcolor[HTML]{C0C0C0}48} & \multirow{-2}{*}{\cellcolor[HTML]{C0C0C0}34}          \\ \hline
\cellcolor[HTML]{FFFFFF}                                    & \cellcolor[HTML]{FFFFFF}                              & \cellcolor[HTML]{FFFFFF}                         & \cellcolor[HTML]{FFFFFF}                         & \cellcolor[HTML]{FFFFFF}                           & \cellcolor[HTML]{FFFFFF}                                  & \cellcolor[HTML]{FFFFFF}                      & \cellcolor[HTML]{FFFFFF}                     & \cellcolor[HTML]{FFFFFF}                     & \cellcolor[HTML]{FFFFFF}                              \\
\multirow{-2}{*}{\cellcolor[HTML]{FFFFFF}\textbf{Circle~\cite{AUTORepo}}}   & \multirow{-2}{*}{\cellcolor[HTML]{FFFFFF}\textbf{71}} & \multirow{-2}{*}{\cellcolor[HTML]{FFFFFF}---}    & \multirow{-2}{*}{\cellcolor[HTML]{FFFFFF}148.63} & \multirow{-2}{*}{\cellcolor[HTML]{FFFFFF}84.37}    & \multirow{-2}{*}{\cellcolor[HTML]{FFFFFF}1078.14}         & \multirow{-2}{*}{\cellcolor[HTML]{FFFFFF}---} & \multirow{-2}{*}{\cellcolor[HTML]{FFFFFF}63} & \multirow{-2}{*}{\cellcolor[HTML]{FFFFFF}71} & \multirow{-2}{*}{\cellcolor[HTML]{FFFFFF}47}          \\ \hline
\cellcolor[HTML]{C0C0C0}                                    & \cellcolor[HTML]{C0C0C0}                              & \cellcolor[HTML]{C0C0C0}                         & \cellcolor[HTML]{C0C0C0}                         & \cellcolor[HTML]{C0C0C0}                           & \cellcolor[HTML]{C0C0C0}                                  & \cellcolor[HTML]{C0C0C0}                      & \cellcolor[HTML]{C0C0C0}                     & \cellcolor[HTML]{C0C0C0}                     & \cellcolor[HTML]{C0C0C0}                              \\
\multirow{-2}{*}{\cellcolor[HTML]{C0C0C0}\textbf{MNIST~\cite{AUTORepo}}}    & \multirow{-2}{*}{\cellcolor[HTML]{C0C0C0}\textbf{38}} & \multirow{-2}{*}{\cellcolor[HTML]{C0C0C0}290.87} & \multirow{-2}{*}{\cellcolor[HTML]{C0C0C0}16.68}  & \multirow{-2}{*}{\cellcolor[HTML]{C0C0C0}4741.53}  & \multirow{-2}{*}{\cellcolor[HTML]{C0C0C0}1265.02}         & \multirow{-2}{*}{\cellcolor[HTML]{C0C0C0}26}  & \multirow{-2}{*}{\cellcolor[HTML]{C0C0C0}38} & \multirow{-2}{*}{\cellcolor[HTML]{C0C0C0}38} & \multirow{-2}{*}{\cellcolor[HTML]{C0C0C0}31}          \\ \hline
\cellcolor[HTML]{FFFFFF}                                    & \cellcolor[HTML]{FFFFFF}                              & \cellcolor[HTML]{FFFFFF}                         & \cellcolor[HTML]{FFFFFF}                         & \cellcolor[HTML]{FFFFFF}                           & \cellcolor[HTML]{FFFFFF}                                  & \cellcolor[HTML]{FFFFFF}                      & \cellcolor[HTML]{FFFFFF}                     & \cellcolor[HTML]{FFFFFF}                     & \cellcolor[HTML]{FFFFFF}                              \\
\multirow{-2}{*}{\cellcolor[HTML]{FFFFFF}\textbf{CIFAR-10~\cite{AUTORepo}}} & \multirow{-2}{*}{\cellcolor[HTML]{FFFFFF}\textbf{46}} & \multirow{-2}{*}{\cellcolor[HTML]{FFFFFF}121.39} & \multirow{-2}{*}{\cellcolor[HTML]{FFFFFF}22.4}   & \multirow{-2}{*}{\cellcolor[HTML]{FFFFFF}10653.63} & \multirow{-2}{*}{\cellcolor[HTML]{FFFFFF}3282.83}         & \multirow{-2}{*}{\cellcolor[HTML]{FFFFFF}46}  & \multirow{-2}{*}{\cellcolor[HTML]{FFFFFF}46} & \multirow{-2}{*}{\cellcolor[HTML]{FFFFFF}46} & \multirow{-2}{*}{\cellcolor[HTML]{FFFFFF}26}          \\ \hline
\end{tabular}}
\label{tab:datasetResult}%

\end{table}

\subsection{\textbf{RQ1 (Validation)} and \textbf{RQ2 (Comparison)}}

Table~\ref{tab:result} and ~\ref{tab:datasetResult} show the evaluation results for RQ1 and RQ2.

\textbf{DeepDiagnosis} (DD) has correctly identified 46 out of 53 buggy models from  the SGS benchmark. DD correctly reported the fault location for 34 models and the failure symptoms for 37 models. Also, DD correctly identified the actionable changes for 28 out 53 faulty models. Lastly, DD identified 138 out of the 203 buggy models from the AUTOTRAINER dataset, correctly reporting fault location, failure symptoms, and actionable changes.



\textbf{DeepLocalize} (DL)~\cite{wardat21DeepLocalize} identified 45 out of the 53 models from the SGS benchmark and indicated fault locations for 26. It reported symptoms for only 23 models, but it cannot provide any suggestions to fix these faults.
Regarding the AUTOTRAINER dataset, DL identified 191 out of the 203 buggy models and correctly reported their fault location. However, DL did not provide any suggestions for fixing those models. Lastly, DL can only detect bugs related to numerical errors. 

\textbf{AUTOTRAINER} (AT)~\cite{ZhangAUTOTRAINER} For the 53 models (SGS benchmark), AT identified 24 buggy models. Out of these, AT successfully reported symptoms for only 8. AT was only able to repair 16 models. DD can handle more varieties of semantically related errors than AT, as shown in Table~\ref{tab:result}. 
Please refer to~\cite{ZhangAUTOTRAINER} for AT's evaluation results while analyzing its dataset.

{
\color{blue}
}

\textbf{UMLUAT} (UM)~\cite{schoop2021umlaut} identified 26 buggy models out of the 53 from the SGS benchmark and found the fault locations for 3. Also, UM reported the symptoms for 17 models and provided the location fix for 15 out of 53. UM correctly identified models and reported possible fix solutions to problems from 72 out of 203 buggy models of the AUTOTRAINER dataset. UM only supports classification problems, while DD supports additional types, such as regression and classification.

\begin{figure}[!htb]
	\centering
	\includegraphics[width=3.3in,clip]{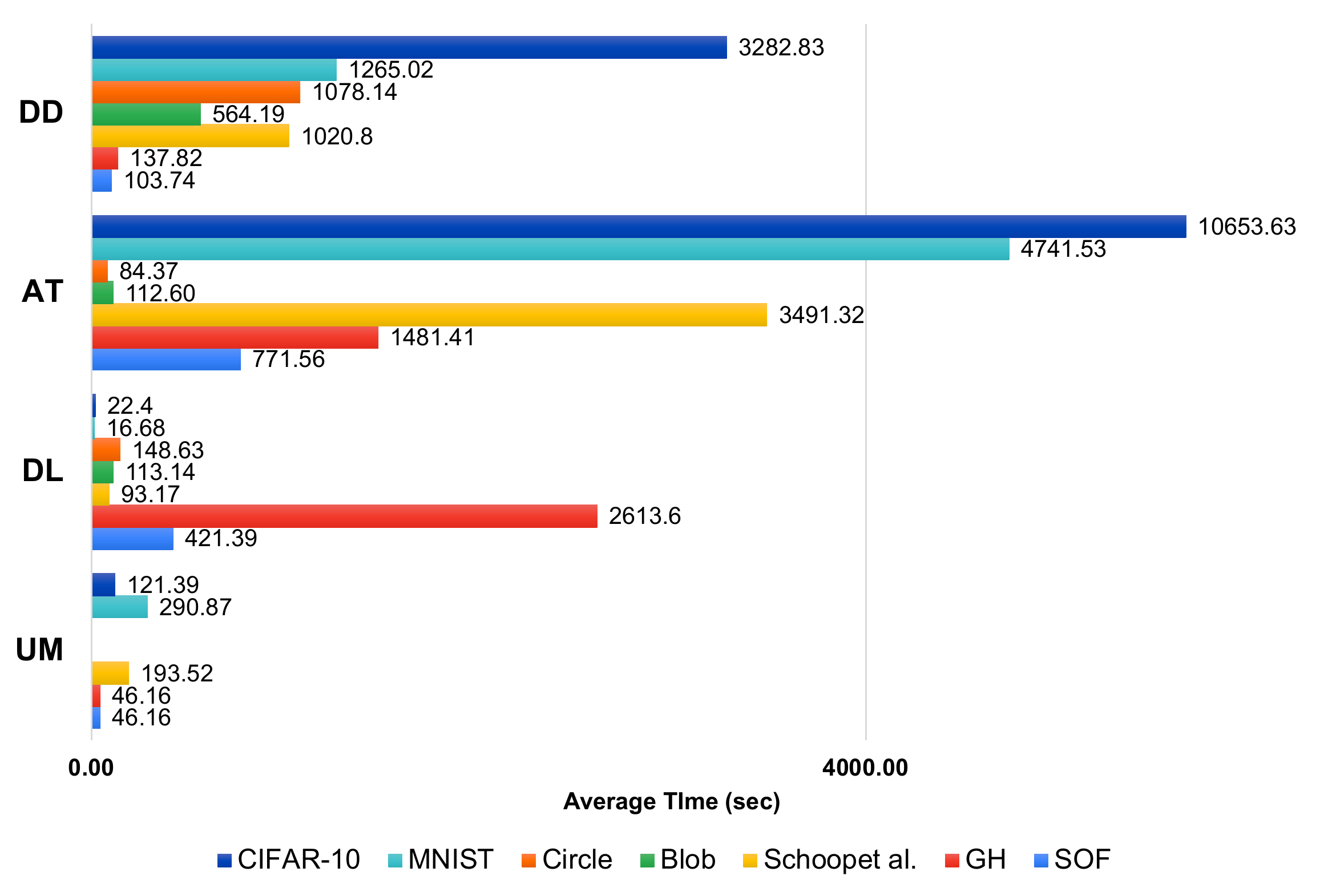}
	\caption{Comparison between UMLUAT (UM) VS DeepLocalize (DL) VS AUTOTRAINER (AT) VS DeepDiagnosis (DD) in terms of seconds}
	\label{Comparsion}
\end{figure}

To evaluate the approaches' overall performance, we collected their total execution time while analyzing the benchmarks. Figure~\ref{Comparsion} shows the results. UM, DL, AT, and DD require on average 46.16, 421.39, 771.56, and 103.74 seconds respectively for all the \sof (SOF) benchmarks. For the \gh (GH) benchmark, the four approaches require on average 46.16, 2613.60, 148.41, and 137.78 seconds respectively. 
For the Schoop~\etal's~\cite{schoop2021umlaut} benchmark, the four approaches take on average 193.52, 93.17, 3491.32, and 1020.80 seconds respectively. For the AUTOTRAINER dataset, the four approaches require on average 4159.25, 4157.36, 170156.70, and 74408.07 seconds respectively to complete their analysis. 
Lastly, the overall average time for UM, DL, AT,  and DD, for all benchmarks is 2972.23, 8388.21,	106490.05, and 44914.17 seconds respectively.

DD' runtime overhead is mainly due to its online dynamic analysis. DD runs its dynamic analysis on the internal parameters of the neural networks, such as the changes of weights and gradients, during the training phase. DD is the most efficient for \sof and Schoop~\etal's model, and is slower than UM on the \gh models. The reason is that DD collects more information than UM during training and checks additional types of error conditions.

DD is faster than AT on all benchmarks except for the Blob and Circle datasets. That is because AT checks the target model after finishing the training phase. DD requires additional time because it validates the model at the end of each epoch during training,
and the number of epochs for these models are between 200 to 500.

\subsection{\textbf{RQ3 (Limitation)}}
Out of 52 programs, our technique failed to identify faults in 6 and localize faults in 18. DD failed to report symptoms for 15 programs and failed to provide the location of fix for 24 (Tables~\ref{tab:addlabel} and ~\ref{tab:result}). In the following, we provide a few examples failed fault localization. 

Our technique does not yet support model with \texttt{fit\_generator()} instead of \texttt{fit()} function.  \texttt{fit\_generator()}  is used for processing a large training dataset that is unable to load into the memory~\cite{kerasDoc}. In the future, we plan to cover more APIs (such as \texttt{fit\_generator()}).

Both \#47 (B3 (C\-10)), and \#53 (B3 (C\-10)) programs are related to checking validation accuracy~\cite{schoop2021umlaut}. The model splits the train data into training and validation data, and then provide the validation data by passing \textit{ validation_data=(x\_val, y\_val)} into the \textit{ fit()} method. The buggy model reported high accuracy for the validation dataset. There may exist an overlap between training data and validation data. But our approach would not indicate any symptom, as it does not support problems related to training and validation.

Both \#43 (A2 (C\-10)), and \#49 (A2 (C\-10)) programs are related to the dropout rate in the \textit{Dropout} layer~\cite{schoop2021umlaut}. The idea of the dropout is to remove certain percentage of neurons during iterations to prevent the overfitting. The buggy model 
sets a high dropout rate = 0.8 which is more than the acceptable rate 50\%. Our approach is not able to provide a correct suggestion to fix the model. In our future work, we plan to investigate more hyperparameters such as the batch size, epoch, and dropout rate to handle the above models.

DD supports deep learning models of various structures, including convolutional neural networks (CNNs) and fully
connected layers. But, Recurrent Neural Networks (RNNs) are not supported by our current reference implementation. Developers can extend our DD to support RNNs and other architectures.

UM only supports classification problems, in which the last layer is \textit{softmax}. Otherwise, it reports false alarms. DL only supports numerical problems, and it does provide any suggestions on how to fix a detected problem. AT supports classification problems and does not support problems in the model architecture (i.e., loss function, activation function at last layer, and some APIs (e.g.,\textit{fit_generator())}). In terms of efficiency, AT takes longer to find a fix, as it tries all possible solutions until arriving at the correct one. In case it does not find an improvement, it marks the problem as unsolvable.


\subsection{\textbf{RQ4 (Ablation)}}

\begin{table}[]
\caption{The Symptoms Results from DeepDiagnosis}
\scalebox{0.60}{
\begin{tabular}{|c|r|r|r|r|r|r|r|r|r|r|}
\hline
                                   & \multicolumn{10}{c|}{\textbf{DD - Symptoms}}                                                                                                                                                                                                                                                                                                                                                                                                                                                          \\ \cline{2-11} 
                                   & \multicolumn{1}{c|}{}                              & \multicolumn{1}{c|}{}                               & \multicolumn{1}{c|}{}                               & \multicolumn{1}{c|}{}                               & \multicolumn{1}{c|}{}                               & \multicolumn{1}{c|}{}                                & \multicolumn{1}{c|}{}                                & \multicolumn{1}{c|}{}                               & \multicolumn{1}{c|}{}         & \multicolumn{1}{c|}{}                          \\
\multirow{-3}{*}{\textbf{Dataset}} & \multicolumn{1}{c|}{\multirow{-2}{*}{\textbf{NS}}} & \multicolumn{1}{c|}{\multirow{-2}{*}{\textbf{UCS}}} & \multicolumn{1}{c|}{\multirow{-2}{*}{\textbf{SAS}}} & \multicolumn{1}{c|}{\multirow{-2}{*}{\textbf{DNS}}} & \multicolumn{1}{c|}{\multirow{-2}{*}{\textbf{ORS}}} & \multicolumn{1}{c|}{\multirow{-2}{*}{\textbf{LNDS}}} & \multicolumn{1}{c|}{\multirow{-2}{*}{\textbf{ANIS}}} & \multicolumn{1}{c|}{\multirow{-2}{*}{\textbf{VGS}}} & \multicolumn{1}{c|}{\multirow{-2}{*}{\textbf{IAS}}} & \multicolumn{1}{c|}{\multirow{-2}{*}{\textbf{ILS}}}\\ \hline
\rowcolor[HTML]{C0C0C0} 
\textbf{\sof~\cite{myRepo}}                       &       15                                             &         1                                            &                      5                               &                 1                                    &     2                                                &                             0                         & 1                                                     &  0                                                   &                               1                &   0  \\ \hline
\textbf{\gh~\cite{myRepo}}                        &       2                                             &        1                                             &                         2                            &                             1                        &   1                                                  &     0                                                 &         0                                             &                    1                                 &            1                                     &  0   \\ \hline
\rowcolor[HTML]{C0C0C0} 
\textbf{Schoop~\etal~\cite{UMLAUTRepo}}                    &   6                                                 &      0                                               &                                             0        &                 3                                    &     2                                                &          0                                            &  0                                                    &          0                                           &       0                                           &   0 \\ \hline
\textbf{Blob~\cite{AUTORepo}}                      & 5                                                  & 7                                                   & 2                                                   & 0                                                   & 0                                                   & 0                                                    & 10                                                   & 10                                                  & 0                                                  & 0 \\ \hline
\rowcolor[HTML]{C0C0C0} 
\textbf{Circle~\cite{AUTORepo}}                    & 12                                                 & 12                                                  & 3                                                   & 1                                                   & 0                                                   & 0                                                    & 11                                                   & 8                                                   & 0                                                  &  0\\ \hline
\textbf{MNIST~\cite{AUTORepo}}                     & 16                                                 & 0                                                   & 0                                                   & 7                                                   & 0                                                   & 0                                                    & 0                                                    & 8                                                   & 0                                                 &   0\\ \hline
\rowcolor[HTML]{C0C0C0} 
\textbf{CIFAR-10~\cite{AUTORepo}}                  & 17                                                 & 4                                                   & 0                                                   & 0                                                   & 0                                                   & 0                                                    & 0                                                    & 5                                                   & 0                                                 &   0\\ \hline
\end{tabular}
}
\label{tab:SymptomsDD}%
\end{table}
The "Ablation" column of Table~\ref{tab:result} shows which procedure in Table~\ref{tab:addlabel} is used to report the symptom in each buggy model for SGS dataset. We found that \textit{ExplodingTensor ()} detects 23 buggy models, \textit{SaturatedActivation ()} detects 7, \textit{DeadNode ()} reports 5,  \textit{OutofRange ()} detects 5, \textit{UnchangeWeight ()} finds 2,   \textit{InvalidAccuracy ()} detects 2,  \textit{AccuracyNotIncreasing ()}, and \textit{VanishingGradient ()} reports only one buggy model.
Table~\ref{tab:SymptomsDD} shows dataset names, and columns contain the number of
symptoms, which were detected successfully by the corresponding procedure in Table~\ref{tab:addlabel}.
From Table~\ref{tab:SymptomsDD}, we found that \textit{ExplodingTensor ()} detects 73 buggy models, \textit{VanishingGradient ()} detects 32,  \textit{UnchangeWeight ()} finds 25, \textit{AccuracyNotIncreasing () 22}, \textit{DeadNode ()} reports 13,  \textit{SaturatedActivation ()} detects 12, \textit{OutofRange ()} detects 5 , and \textit{InvalidAccuracy ()} reports only two buggy models. Although the incorrect DNN models related to parameters and structures often manifest as numerical errors during training, DD provided further reasoning and categories of causes using these procedures, which can help quickly fix the bugs. Our study also found that data preparation is a frequently occurring issue and thus the \textit{ImproperData()} procedure is frequently invoked. SGS benchmark does not have a very deep model that contains many layers. Thus we did not use \textit{VanishingGradient ()} detector very frequently. On the other hand, \textit{VanishingGradient ()} is invoked very frequently in AUTOTRAINER models, because this dataset has many layers using sigmoid and tanh as activation functions. However, when N layers use a Logistic activation function (like sigmoid or tanh), N small derivatives are multiplied together. Thus, the gradient decreases exponentially and propagates down to the input layer.

\subsection{\textbf{Results Discussions}}
We compared and contrasted three approaches~\cite{schoop2020scram,wardat21DeepLocalize,ZhangAUTOTRAINER} against our approach (DD). From Table~\ref{tab:datasetResult}, we found our approach detected more problems in the SGS dataset than AUTOTRAINER. Also, it detected fewer problems in AUTOTRAINER dataset than the AT approach. The reason is that our approach only reported the problem and solution if it detected one of 8 symptoms. On the other hand, AT inspects the model based on the training accuracy threshold~\cite{ZhangAUTOTRAINER}.

For our evaluation, we used 188 normal models from~\cite{ZhangAUTOTRAINER}.  78 are MNIST, 35 are CIFAR-10, 36 are Circle, and 39 are Blob. UM reported the message: ``<Warning: Possible overfitting>'' for 68 out 78 MNIST models. It reported the following message: ``[<Error: Input data exceeds typical limits>]'' for 35 out 35 CIFAR-10 models, because the training data is not in the range $[$-1, 1$]$. DL reported message: ``MDL: Model Does not Learn'' for 4 out 34 Circle models and 16 out 39 Blob models. For all MNIST and CIFAR-10 models, DL reported different messages. AT checks if a model has training accuracy less than or equal to the threshold of 60\%. To make a fair comparison between the approaches, we changed the training accuracy threshold to 100\%. AT reported different symptoms for 10 out of 36 Circle, 5 out of 39 Blob models, and 2 models with problems out of the 78 MNIST models. Our approach reported one saturated symptom for 36 Circle, which is not supported in AT, reported 8 symptoms - 6 ``saturated activations'' and 2 the ``accuracy is not increasing.'' For the MNIST model, our approach reported 37 symptoms - 35 ``dead nodes'' and one is a ``numerical problem;'' we investigated this model and found its accuracy is 20\%. For CIFAR-10 models, DD reported 21 models with ``dead node'' out of 35 models. All detailed experiment results are publicly available~\cite{DDRepo}.

\subsection{\textbf{Summary}}
DD significantly outperformed the baselines UM, DL, and AT in the SGS dataset (Tables~\ref{tab:result} and~\ref{tab:datasetResult}). In particular, identified 46 out of 53  buggy models, correctly performed fault localization in 34 models, and reported symptoms for 37 of those.  DD also provided a location to fix 28 out of 53 faulty models. Regarding total analysis time, DD outperformed DL and AT in benchmarks. As DD does not require the training phase to finish to detect bugs. Also, DD uses a Decision Tree (Figure~\ref{DecisionTable}) approach to reduce the search space when mapping symptoms to their root causes.

Furthermore, DD is more comprehensive than prior work, as it supports several varieties and semantically related errors in classification and regression models. Also, DD supports 8 failure symptoms, while prior approaches support fewer (in Section~\ref{sec:Approach}).

Finally, DD does not support some APIs (e.g., \textit{fit\_generator()}) as we consider problems related to hyperparameters, for example, epoch, batch size, and dropout rate, as out of scope. 






\section[Threats]{Threats to Validity}
\label{sec:THREATSTOVALIDITY}
\textbf{\textit{External Threat:}} We have collected 53 real-world buggy DNN models from \sof, \gh and 496 models from prior work~\cite{wardat21DeepLocalize,schoop2021umlaut, ZhangAUTOTRAINER}. These models cover a variety of failure symptoms and location to perform fixes; however, our dataset may not include all types of DNN APIs and their parameters. To mitigate the threat of behavior changes caused by the Extractor tool, used to extract the source code from the 496 models~\cite{ZhangAUTOTRAINER}. We have verified the accuracy of each model before and after their conversion. In terms of execution time, different hardware configurations may offer varying response times. We mitigated this threat by executing our experiments several times and calculated their averages.



\textbf{\textit{Internal Threat:}} When implementing Algorithm~\ref{alg:fault-localize}, Decision Tree (Figure~\ref{DecisionTable}), and Tables~\ref{tab:addlabel} and~\ref{tab:checker},  
we used the parameters defined by prior works~\cite{SageMaker, LearningRate, schoop2020scram, goodfellow2016deep, ZhangAUTOTRAINER}. 
These selected values may not work for some unseen examples. To mitigate this threat, we have validated these selected parameters against our benchmarks collected from a diverse set of sources~\cite{wardat21DeepLocalize,schoop2021umlaut, ZhangAUTOTRAINER}. 
For each of these benchmarks, our selected parameters work consistently well. 
Although we have carefully inspected our code, our implementation may still contain some errors.
We manually constructed ground truths regarding fault location, failure symptoms, and location to fix for all the buggy models based on the data from the previous research~\cite{wardat21DeepLocalize, schoop2021umlaut, ZhangAUTOTRAINER}. 
This process may have introduced errors.

\section[Related]{Related Work}
\label{sec:relatedwork}



\textit{\textbf{Fault localization in Deep Neural Networks:}} The recent increase in the popularity of deep learning apps has motivated researchers to adapt fault localization techniques to this context. With the intent of validating different parts of DL-based systems and discovering faulty behaviors. The goal of fault localization is to identify suspicious methods and statements, to isolate the root causes of program failures, and reduce the effort of fixing the fault~\cite{pearson2017evaluating}.
Wardat~\etal~\cite{wardat21DeepLocalize} presented an automatic approach for fault localization called DeepLocalize. It performs dynamic analysis during training to determines if a target model contains any bugs. It identifies the root causes by catching numerical errors during DNN training. While DeepLocalize focuses on identifying bugs and faults based on numerical errors, DeepDiagnosis aims to perform fault localization beyond that scope. Furthermore, our approach can report symptoms and provide actionable fixes to a problem.



DEBAR~\cite{zhang2020detecting} is a static analysis approach that detects numerical bugs in DNNs. DEBAR uses two abstraction techniques to improve its precision and scalability. DeepDiagnosis uses dynamic analysis to localize faults and report symptoms of a model during training. In contrast, DEBAR uses a static analysis approach to detect numerical bugs with two abstraction techniques. 

Schoop~\etal~\cite{schoop2021umlaut} proposed UMLUAT, a user interface tool to find, understand and fix deep learning bugs using heuristics. It enables users to check the structure of DNN programs and model behavior during training. Then, it provides readable error messages to assist users in understanding and fixing bugs. Section~\secref{sec:EVALUATION} shows the comparison between UMLUAT~\cite{schoop2021umlaut} and DeepDiagnosis. DeepDiagnosis is more comprehensive, efficient, and effective than UMLAUT, which only supports classification models.

DeepFault~\cite{DeepFault} is an approach that identifies suspicious neurons of a DNN and then fixes these errors by generating samples for retraining the model. DeepFault is inspired by spectrum-based fault localization. It counts the number of times a neuron was active/inactive when the network made a successful or failed decision. It then calculates a suspiciousness score such as the spectrum-based fault localization tool Tarantula. In contrast, DeepDiagnosis focuses on identifying faults and reporting different types of symptoms for structure bugs.

\textit{\textbf{Bug Repair in Deep Neural Networks:}}
Zhang~\etal~\cite{zhang2019apricot} proposed Apricot, an approach for automatically repairing deep learning models. Apricot aims to fix ill-trained weights without requiring additional training data or any artificial parameters in the DNN. MODE~\cite{ma2018mode} is a white-box approach that focuses on improving the model performance. It is an automated debugging technique inspired by state differential analysis. MODE can determine whether a model has overfitting or under-fitting problems. Compared with MODE and Apricot, which focus on training bugs (e.g., insufficient training data), DeepDiagnosis focuses on structure bugs (e.g., activation function misused).

Zhang~\etal~\cite{ZhangAUTOTRAINER} introduced AUTOTRAINER, an approach for fixing classification problems. Zhang~\etal define five symptoms, and provide a set of possible solutions to fix each one. Once AUTOTRAINER detects a problem, it tries the candidate solutions, one by one, until it addresses the problem. If none of the solutions fix the problem, it reports a failure message.  
The evaluation used six popular datasets and showed that AUTOTRAINER detects and repairs the models based on a specific threshold. AUTOTRAINER was able to improve the accuracy for all repairing models on average 47.08\%. DeepDiagnosis analyzes the model's source code during the training phase to localize the bug. DeepDiagnosis supports eight symptoms, while AUTOTRAINER supports five. DeepDiagnosis does not perform automated fixes, but it provides actionable recommendations that developers can follow. AUTOTRAINER tries all possible strategies in its search space to fix a problem and outputs whether or not the fix was successful. In contrast, DeepDiagnosis uses a decision tree to reduce the solution search space, thus saving time and computational resources.
In summary, the goals of DeepDiagnosis and AUTOTRAINER are different; DeepDiagnosis focuses on fault localization while AUTOTRAINER on automatically repairing a model.

\section[Conclusion]{Conclusions and Future Work}
\label{sec:conclusion}
This paper introduces a dynamic analysis approach called DeepDiagnosis that a non-expert can use to detect errors and receive useful messages for diagnosing and fixing the DNN models. DeepDiagnosis provides a list of verification procedures to automatically detect 8 types of common symptoms.  Our results show that DeepDiagnosis can successfully detect different types of symptoms and report actionable changes. It outperforms the state of the art tool such as UMLUAT and DeepLocalize, and it is faster than AUTOTRAINER for fault localization and provide suggestions to fix the issue. In the future, we plan to expand this prototype to handle more types of models  and failure symptoms, and also automatically fix the bugs.




\balance
\bibliographystyle{ACM-Reference-Format}
\bibliography{refs}


\begin{thebibliography}{44}


\ifx \showCODEN    \undefined \def \showCODEN     #1{\unskip}     \fi
\ifx \showDOI      \undefined \def \showDOI       #1{#1}\fi
\ifx \showISBNx    \undefined \def \showISBNx     #1{\unskip}     \fi
\ifx \showISBNxiii \undefined \def \showISBNxiii  #1{\unskip}     \fi
\ifx \showISSN     \undefined \def \showISSN      #1{\unskip}     \fi
\ifx \showLCCN     \undefined \def \showLCCN      #1{\unskip}     \fi
\ifx \shownote     \undefined \def \shownote      #1{#1}          \fi
\ifx \showarticletitle \undefined \def \showarticletitle #1{#1}   \fi
\ifx \showURL      \undefined \def \showURL       {\relax}        \fi
\providecommand\bibfield[2]{#2}
\providecommand\bibinfo[2]{#2}
\providecommand\natexlab[1]{#1}
\providecommand\showeprint[2][]{arXiv:#2}

\bibitem[\protect\citeauthoryear{??}{Lea}{2015}]%
        {LearningRate}
 \bibinfo{year}{2015}\natexlab{}.
\newblock
  \bibinfo{howpublished}{\url{https://cs231n.github.io/neural-networks-3/}}.
\newblock
\newblock
\shownote{[Online; accessed 20-Aug-2020].}


\bibitem[\protect\citeauthoryear{??}{bug}{2016}]%
        {bug10}
 \bibinfo{year}{2016}\natexlab{}.
\newblock \bibinfo{title}{{How to prepare a dataset for Keras?}}
\newblock
  \bibinfo{howpublished}{\url{https://stackoverflow.com/questions/31880720/}}.
\newblock
\newblock
\shownote{[Online; accessed 19-Aug-2020].}


\bibitem[\protect\citeauthoryear{{}}{{}}{2020a}]%
        {Manifold}
\bibfield{author}{\bibinfo{person}{{}}.} \bibinfo{year}{2020}\natexlab{a}.
\newblock \bibinfo{title}{{Manifold}}.
\newblock
\newblock
\newblock
\shownote{\url{https://github.com/uber/manifold}.}


\bibitem[\protect\citeauthoryear{{}}{{}}{2020b}]%
        {TensorWatch}
\bibfield{author}{\bibinfo{person}{{}}.} \bibinfo{year}{2020}\natexlab{b}.
\newblock \bibinfo{title}{{Tensorwatch}}.
\newblock
\newblock
\newblock
\shownote{\url{https://github.com/microsoft/tensorwatch}.}


\bibitem[\protect\citeauthoryear{{}}{{}}{2020c}]%
        {Visdom}
\bibfield{author}{\bibinfo{person}{{}}.} \bibinfo{year}{2020}\natexlab{c}.
\newblock \bibinfo{title}{{Visdom}}.
\newblock
\newblock
\newblock
\shownote{\url{https://github.com/fossasia/visdom}.}


\bibitem[\protect\citeauthoryear{??}{DDR}{2021}]%
        {DDRepo}
 \bibinfo{year}{2021}\natexlab{}.
\newblock
  \bibinfo{howpublished}{\url{https://github.com/DeepDiagnosis/ICSE2022}}.
\newblock
\newblock
\shownote{[Online; accessed 12-August-2021].}


\bibitem[\protect\citeauthoryear{??}{myR}{2021}]%
        {myRepo}
 \bibinfo{year}{2021}\natexlab{}.
\newblock
  \bibinfo{howpublished}{\url{https://github.com/Wardat-ISU/DeepLocalize}}.
\newblock
\newblock
\shownote{[Online; accessed 12-Aug-2021].}


\bibitem[\protect\citeauthoryear{??}{UML}{2021}]%
        {UMLAUTRepo}
 \bibinfo{year}{2021}\natexlab{}.
\newblock \bibinfo{howpublished}{\url{https://github.com/BerkeleyHCI/umlaut}}.
\newblock
\newblock
\shownote{[Online; accessed 12-Aug-2021].}


\bibitem[\protect\citeauthoryear{??}{AUT}{2021}]%
        {AUTORepo}
 \bibinfo{year}{2021}\natexlab{}.
\newblock
  \bibinfo{howpublished}{\url{https://github.com/shiningrain/AUTOTRAINER}}.
\newblock
\newblock
\shownote{[Online; accessed 12-August-2021].}


\bibitem[\protect\citeauthoryear{{Amazon}}{{Amazon}}{2017}]%
        {SageMaker}
\bibfield{author}{\bibinfo{person}{{Amazon}}.} \bibinfo{year}{2017}\natexlab{}.
\newblock \bibinfo{title}{{Amazon SageMaker}}.
\newblock
\newblock
\newblock
\shownote{\url{https://docs.aws.amazon.com/sagemaker/latest/dg/whatis.html}.}


\bibitem[\protect\citeauthoryear{Eniser, Gerasimou, and Sen}{Eniser
  et~al\mbox{.}}{2019}]%
        {DeepFault}
\bibfield{author}{\bibinfo{person}{Hasan~Ferit Eniser}, \bibinfo{person}{Simos
  Gerasimou}, {and} \bibinfo{person}{Alper Sen}.}
  \bibinfo{year}{2019}\natexlab{}.
\newblock \showarticletitle{{DeepFault}: fault localization for deep neural
  networks}. In \bibinfo{booktitle}{\emph{Fundamental Approaches to Software
  Engineering}}, \bibfield{editor}{\bibinfo{person}{Reiner H{\"a}hnle} {and}
  \bibinfo{person}{Wil van~der Aalst}} (Eds.). \bibinfo{publisher}{Springer
  International Publishing}, \bibinfo{address}{Cham},
  \bibinfo{pages}{171--191}.
\newblock
\showISBNx{978-3-030-16722-6}


\bibitem[\protect\citeauthoryear{Evci}{Evci}{2018}]%
        {evci2018detecting}
\bibfield{author}{\bibinfo{person}{Utku Evci}.}
  \bibinfo{year}{2018}\natexlab{}.
\newblock \showarticletitle{Detecting dead weights and units in neural
  networks}.
\newblock \bibinfo{journal}{\emph{arXiv preprint arXiv:1806.06068}}
  (\bibinfo{year}{2018}).
\newblock


\bibitem[\protect\citeauthoryear{{Francois Chollet}}{{Francois
  Chollet}}{2015a}]%
        {kerasDoc}
\bibfield{author}{\bibinfo{person}{{Francois Chollet}}.}
  \bibinfo{year}{2015}\natexlab{a}.
\newblock \bibinfo{title}{{Keras documentation}}.
\newblock
\newblock
\newblock
\shownote{\url{https://keras.io/}.}


\bibitem[\protect\citeauthoryear{{Francois Chollet}}{{Francois
  Chollet}}{2015b}]%
        {Keras}
\bibfield{author}{\bibinfo{person}{{Francois Chollet}}.}
  \bibinfo{year}{2015}\natexlab{b}.
\newblock \bibinfo{title}{Keras: the {Python} Deep Learning library}.
\newblock
\newblock
\newblock
\shownote{\url{https://keras.io/}.}


\bibitem[\protect\citeauthoryear{Glorot and Bengio}{Glorot and Bengio}{2010}]%
        {glorot2010understanding}
\bibfield{author}{\bibinfo{person}{Xavier Glorot} {and} \bibinfo{person}{Yoshua
  Bengio}.} \bibinfo{year}{2010}\natexlab{}.
\newblock \showarticletitle{Understanding the difficulty of training deep
  feedforward neural networks}. In \bibinfo{booktitle}{\emph{Proceedings of the
  thirteenth international conference on artificial intelligence and
  statistics}}. JMLR Workshop and Conference Proceedings,
  \bibinfo{pages}{249--256}.
\newblock


\bibitem[\protect\citeauthoryear{Goodfellow, Bengio, Courville, and
  Bengio}{Goodfellow et~al\mbox{.}}{2016}]%
        {goodfellow2016deep}
\bibfield{author}{\bibinfo{person}{Ian Goodfellow}, \bibinfo{person}{Yoshua
  Bengio}, \bibinfo{person}{Aaron Courville}, {and} \bibinfo{person}{Yoshua
  Bengio}.} \bibinfo{year}{2016}\natexlab{}.
\newblock \bibinfo{booktitle}{\emph{Deep learning}}. Vol.~\bibinfo{volume}{1}.
\newblock \bibinfo{publisher}{MIT press Cambridge}.
\newblock


\bibitem[\protect\citeauthoryear{Grossman, Fitzmaurice, and Attar}{Grossman
  et~al\mbox{.}}{2009}]%
        {grossman2009survey}
\bibfield{author}{\bibinfo{person}{Tovi Grossman}, \bibinfo{person}{George
  Fitzmaurice}, {and} \bibinfo{person}{Ramtin Attar}.}
  \bibinfo{year}{2009}\natexlab{}.
\newblock \showarticletitle{A survey of software learnability: metrics,
  methodologies and guidelines}. In \bibinfo{booktitle}{\emph{Proceedings of
  the sigchi conference on human factors in computing systems}}.
  \bibinfo{pages}{649--658}.
\newblock


\bibitem[\protect\citeauthoryear{{Guido van Rossum}}{{Guido van
  Rossum}}{2019}]%
        {difflib}
\bibfield{author}{\bibinfo{person}{{Guido van Rossum}}.}
  \bibinfo{year}{2019}\natexlab{}.
\newblock \bibinfo{title}{{Module difflib}}.
\newblock
\newblock
\newblock
\shownote{\url{https://github.com/python/cpython/blob/3.9/Lib/difflib.py}.}


\bibitem[\protect\citeauthoryear{Gulcehre, Moczulski, Denil, and
  Bengio}{Gulcehre et~al\mbox{.}}{2016}]%
        {gulcehre2016noisy}
\bibfield{author}{\bibinfo{person}{Caglar Gulcehre}, \bibinfo{person}{Marcin
  Moczulski}, \bibinfo{person}{Misha Denil}, {and} \bibinfo{person}{Yoshua
  Bengio}.} \bibinfo{year}{2016}\natexlab{}.
\newblock \showarticletitle{Noisy activation functions}. In
  \bibinfo{booktitle}{\emph{International conference on machine learning}}.
  PMLR, \bibinfo{pages}{3059--3068}.
\newblock


\bibitem[\protect\citeauthoryear{Isac, J~Frederico, Kragic, and Stork}{Isac
  et~al\mbox{.}}{2020}]%
        {isac2020effect}
\bibfield{author}{\bibinfo{person}{Arnekvist Isac}, \bibinfo{person}{Carvalho
  J~Frederico}, \bibinfo{person}{Danica Kragic}, {and}
  \bibinfo{person}{Johannes~Andreas Stork}.} \bibinfo{year}{2020}\natexlab{}.
\newblock \showarticletitle{The effect of Target Normalization and Momentum on
  Dying ReLU}. In \bibinfo{booktitle}{\emph{The 32nd annual workshop of the
  Swedish Artificial Intelligence Society (SAIS)}}.
\newblock


\bibitem[\protect\citeauthoryear{Islam, Nguyen, Pan, and Rajan}{Islam
  et~al\mbox{.}}{2019}]%
        {islam19}
\bibfield{author}{\bibinfo{person}{Md~Johirul Islam}, \bibinfo{person}{Giang
  Nguyen}, \bibinfo{person}{Rangeet Pan}, {and} \bibinfo{person}{Hridesh
  Rajan}.} \bibinfo{year}{2019}\natexlab{}.
\newblock \showarticletitle{A Comprehensive Study on Deep Learning Bug
  Characteristics}. In \bibinfo{booktitle}{\emph{ESEC/FSE'19: The ACM Joint
  European Software Engineering Conference and Symposium on the Foundations of
  Software Engineering (ESEC/FSE)}} \emph{(\bibinfo{series}{ESEC/FSE 2019})}.
\newblock


\bibitem[\protect\citeauthoryear{Islam, Pan, Nguyen, and Rajan}{Islam
  et~al\mbox{.}}{2020}]%
        {islam20repairing}
\bibfield{author}{\bibinfo{person}{Md~Johirul Islam}, \bibinfo{person}{Rangeet
  Pan}, \bibinfo{person}{Giang Nguyen}, {and} \bibinfo{person}{Hridesh Rajan}.}
  \bibinfo{year}{2020}\natexlab{}.
\newblock \showarticletitle{Repairing Deep Neural Networks: Fix Patterns and
  Challenges}. In \bibinfo{booktitle}{\emph{ICSE'20: The 42nd International
  Conference on Software Engineering}} (Seoul, South Korea).
\newblock


\bibitem[\protect\citeauthoryear{Janowczyk and Madabhushi}{Janowczyk and
  Madabhushi}{2016}]%
        {janowczyk2016deep}
\bibfield{author}{\bibinfo{person}{Andrew Janowczyk} {and}
  \bibinfo{person}{Anant Madabhushi}.} \bibinfo{year}{2016}\natexlab{}.
\newblock \showarticletitle{Deep learning for digital pathology image analysis:
  A comprehensive tutorial with selected use cases}.
\newblock \bibinfo{journal}{\emph{Journal of pathology informatics}}
  \bibinfo{volume}{7} (\bibinfo{year}{2016}).
\newblock


\bibitem[\protect\citeauthoryear{Knox}{Knox}{2018}]%
        {knox2018machine}
\bibfield{author}{\bibinfo{person}{Steven~W Knox}.}
  \bibinfo{year}{2018}\natexlab{}.
\newblock \bibinfo{booktitle}{\emph{Machine learning: a concise introduction}}.
  Vol.~\bibinfo{volume}{285}.
\newblock \bibinfo{publisher}{John Wiley \& Sons}.
\newblock


\bibitem[\protect\citeauthoryear{Kong and Takatsuka}{Kong and
  Takatsuka}{2017}]%
        {kong2017hexpo}
\bibfield{author}{\bibinfo{person}{Shumin Kong} {and} \bibinfo{person}{Masahiro
  Takatsuka}.} \bibinfo{year}{2017}\natexlab{}.
\newblock \showarticletitle{Hexpo: A vanishing-proof activation function}. In
  \bibinfo{booktitle}{\emph{2017 International Joint Conference on Neural
  Networks (IJCNN)}}. IEEE, \bibinfo{pages}{2562--2567}.
\newblock


\bibitem[\protect\citeauthoryear{Ma, Liu, Lee, Zhang, and Grama}{Ma
  et~al\mbox{.}}{2018}]%
        {ma2018mode}
\bibfield{author}{\bibinfo{person}{Shiqing Ma}, \bibinfo{person}{Yingqi Liu},
  \bibinfo{person}{Wen-Chuan Lee}, \bibinfo{person}{Xiangyu Zhang}, {and}
  \bibinfo{person}{Ananth Grama}.} \bibinfo{year}{2018}\natexlab{}.
\newblock \showarticletitle{MODE: automated neural network model debugging via
  state differential analysis and input selection}. In
  \bibinfo{booktitle}{\emph{Proceedings of the 2018 26th ACM Joint Meeting on
  European Software Engineering Conference and Symposium on the Foundations of
  Software Engineering}}. \bibinfo{pages}{175--186}.
\newblock


\bibitem[\protect\citeauthoryear{Man{\'e} et~al\mbox{.}}{Man{\'e}
  et~al\mbox{.}}{2015}]%
        {mane2015tensorboard}
\bibfield{author}{\bibinfo{person}{D Man{\'e}} {et~al\mbox{.}}}
  \bibinfo{year}{2015}\natexlab{}.
\newblock \bibinfo{title}{TensorBoard: TensorFlow’s visualization toolkit}.
\newblock
\newblock


\bibitem[\protect\citeauthoryear{{Martin Abadi et al}}{{Martin Abadi et
  al}}{2015}]%
        {TensorFlow}
\bibfield{author}{\bibinfo{person}{{Martin Abadi et al}}.}
  \bibinfo{year}{2015}\natexlab{}.
\newblock \bibinfo{title}{{TensorFlow}: large-Scale Machine Learning on
  Heterogeneous Systems}.
\newblock
\newblock
\newblock
\shownote{\url{https://www.tensorflow.org/}.}


\bibitem[\protect\citeauthoryear{Miller and Hardt}{Miller and Hardt}{2018}]%
        {miller2018stable}
\bibfield{author}{\bibinfo{person}{John Miller} {and} \bibinfo{person}{Moritz
  Hardt}.} \bibinfo{year}{2018}\natexlab{}.
\newblock \showarticletitle{Stable recurrent models}.
\newblock \bibinfo{journal}{\emph{arXiv preprint arXiv:1805.10369}}
  (\bibinfo{year}{2018}).
\newblock


\bibitem[\protect\citeauthoryear{Miotto, Wang, Wang, Jiang, and Dudley}{Miotto
  et~al\mbox{.}}{2018}]%
        {miotto2018deep}
\bibfield{author}{\bibinfo{person}{Riccardo Miotto}, \bibinfo{person}{Fei
  Wang}, \bibinfo{person}{Shuang Wang}, \bibinfo{person}{Xiaoqian Jiang}, {and}
  \bibinfo{person}{Joel~T Dudley}.} \bibinfo{year}{2018}\natexlab{}.
\newblock \showarticletitle{Deep learning for healthcare: review, opportunities
  and challenges}.
\newblock \bibinfo{journal}{\emph{Briefings in bioinformatics}}
  \bibinfo{volume}{19}, \bibinfo{number}{6} (\bibinfo{year}{2018}),
  \bibinfo{pages}{1236--1246}.
\newblock


\bibitem[\protect\citeauthoryear{Pearson, Campos, Just, Fraser, Abreu, Ernst,
  Pang, and Keller}{Pearson et~al\mbox{.}}{2017}]%
        {pearson2017evaluating}
\bibfield{author}{\bibinfo{person}{Spencer Pearson}, \bibinfo{person}{Jos{\'e}
  Campos}, \bibinfo{person}{Ren{\'e} Just}, \bibinfo{person}{Gordon Fraser},
  \bibinfo{person}{Rui Abreu}, \bibinfo{person}{Michael~D Ernst},
  \bibinfo{person}{Deric Pang}, {and} \bibinfo{person}{Benjamin Keller}.}
  \bibinfo{year}{2017}\natexlab{}.
\newblock \showarticletitle{Evaluating and improving fault localization}. In
  \bibinfo{booktitle}{\emph{2017 IEEE/ACM 39th International Conference on
  Software Engineering (ICSE)}}. IEEE, \bibinfo{pages}{609--620}.
\newblock


\bibitem[\protect\citeauthoryear{Schoop, Huang, and Hartmann}{Schoop
  et~al\mbox{.}}{2020}]%
        {schoop2020scram}
\bibfield{author}{\bibinfo{person}{Eldon Schoop}, \bibinfo{person}{Forrest
  Huang}, {and} \bibinfo{person}{Bj{\"o}rn Hartmann}.}
  \bibinfo{year}{2020}\natexlab{}.
\newblock \showarticletitle{SCRAM: Simple Checks for Realtime Analysis of Model
  Training for Non-Expert ML Programmers}. In
  \bibinfo{booktitle}{\emph{Extended Abstracts of the 2020 CHI Conference on
  Human Factors in Computing Systems}}. \bibinfo{pages}{1--10}.
\newblock


\bibitem[\protect\citeauthoryear{Schoop, Huang, and Hartmann}{Schoop
  et~al\mbox{.}}{2021}]%
        {schoop2021umlaut}
\bibfield{author}{\bibinfo{person}{Eldon Schoop}, \bibinfo{person}{Forrest
  Huang}, {and} \bibinfo{person}{Bj{\"o}rn Hartmann}.}
  \bibinfo{year}{2021}\natexlab{}.
\newblock \showarticletitle{UMLAUT: Debugging Deep Learning Programs using
  Program Structure and Model Behavior}. In
  \bibinfo{booktitle}{\emph{Proceedings of the 2021 CHI Conference Extended
  Abstracts on Human Factors in Computing Systems}}.
\newblock


\bibitem[\protect\citeauthoryear{{Shanqing Cai}}{{Shanqing Cai}}{2017}]%
        {tfdbg}
\bibfield{author}{\bibinfo{person}{{Shanqing Cai}}.}
  \bibinfo{year}{2017}\natexlab{}.
\newblock \bibinfo{title}{{Debug TensorFlow Models with tfdbg}}.
\newblock
\newblock
\newblock
\shownote{\url{https://developers.googleblog.com/2017/02/debug-tensorflow-models-with-tfdbg.html}.}


\bibitem[\protect\citeauthoryear{Sussillo and Abbott}{Sussillo and
  Abbott}{2014}]%
        {sussillo2014random}
\bibfield{author}{\bibinfo{person}{David Sussillo} {and} \bibinfo{person}{LF
  Abbott}.} \bibinfo{year}{2014}\natexlab{}.
\newblock \showarticletitle{Random walk initialization for training very deep
  feedforward networks}.
\newblock \bibinfo{journal}{\emph{arXiv preprint arXiv:1412.6558}}
  (\bibinfo{year}{2014}).
\newblock


\bibitem[\protect\citeauthoryear{Tan and Lim}{Tan and Lim}{2019}]%
        {tan2019vanishing}
\bibfield{author}{\bibinfo{person}{Hong~Hui Tan} {and}
  \bibinfo{person}{King~Hann Lim}.} \bibinfo{year}{2019}\natexlab{}.
\newblock \showarticletitle{Vanishing gradient mitigation with deep learning
  neural network optimization}. In \bibinfo{booktitle}{\emph{2019 7th
  International Conference on Smart Computing \& Communications (ICSCC)}}.
  IEEE, \bibinfo{pages}{1--4}.
\newblock


\bibitem[\protect\citeauthoryear{Tian, Pei, Jana, and Ray}{Tian
  et~al\mbox{.}}{2018}]%
        {DeepTest}
\bibfield{author}{\bibinfo{person}{Yuchi Tian}, \bibinfo{person}{Kexin Pei},
  \bibinfo{person}{Suman Jana}, {and} \bibinfo{person}{Baishakhi Ray}.}
  \bibinfo{year}{2018}\natexlab{}.
\newblock \showarticletitle{{DeepTest}: automated Testing of
  Deep-Neural-Network-Driven Autonomous Cars}. In
  \bibinfo{booktitle}{\emph{Proceedings of the 40th International Conference on
  Software Engineering}} (Gothenburg, Sweden) \emph{(\bibinfo{series}{ICSE
  ’18})}. \bibinfo{publisher}{Association for Computing Machinery},
  \bibinfo{address}{New York, NY, USA}, \bibinfo{pages}{303–314}.
\newblock
\showISBNx{9781450356381}
\urldef\tempurl%
\url{https://doi.org/10.1145/3180155.3180220}
\showDOI{\tempurl}


\bibitem[\protect\citeauthoryear{Veres and Moussa}{Veres and Moussa}{2019}]%
        {veres2019deep}
\bibfield{author}{\bibinfo{person}{Matthew Veres} {and} \bibinfo{person}{Medhat
  Moussa}.} \bibinfo{year}{2019}\natexlab{}.
\newblock \showarticletitle{Deep learning for intelligent transportation
  systems: A survey of emerging trends}.
\newblock \bibinfo{journal}{\emph{IEEE Transactions on Intelligent
  transportation systems}} \bibinfo{volume}{21}, \bibinfo{number}{8}
  (\bibinfo{year}{2019}), \bibinfo{pages}{3152--3168}.
\newblock


\bibitem[\protect\citeauthoryear{Wardat, Le, and Rajan}{Wardat
  et~al\mbox{.}}{2021}]%
        {wardat21DeepLocalize}
\bibfield{author}{\bibinfo{person}{Mohammad Wardat}, \bibinfo{person}{Wei Le},
  {and} \bibinfo{person}{Hridesh Rajan}.} \bibinfo{year}{2021}\natexlab{}.
\newblock \showarticletitle{DeepLocalize: fault localization for deep neural
  networks}. In \bibinfo{booktitle}{\emph{ICSE'21: The 43nd International
  Conference on Software Engineering}}.
\newblock


\bibitem[\protect\citeauthoryear{Xu, Huang, and Li}{Xu et~al\mbox{.}}{2016}]%
        {xu2016revise}
\bibfield{author}{\bibinfo{person}{Bing Xu}, \bibinfo{person}{Ruitong Huang},
  {and} \bibinfo{person}{Mu Li}.} \bibinfo{year}{2016}\natexlab{}.
\newblock \showarticletitle{Revise saturated activation functions}.
\newblock \bibinfo{journal}{\emph{arXiv preprint arXiv:1602.05980}}
  (\bibinfo{year}{2016}).
\newblock


\bibitem[\protect\citeauthoryear{Zhang and Chan}{Zhang and Chan}{2019}]%
        {zhang2019apricot}
\bibfield{author}{\bibinfo{person}{Hao Zhang} {and} \bibinfo{person}{WK Chan}.}
  \bibinfo{year}{2019}\natexlab{}.
\newblock \showarticletitle{Apricot: a weight-adaptation approach to fixing
  deep learning models}. In \bibinfo{booktitle}{\emph{2019 34th IEEE/ACM
  International Conference on Automated Software Engineering (ASE)}}. IEEE,
  \bibinfo{pages}{376--387}.
\newblock


\bibitem[\protect\citeauthoryear{Zhang, Gupta, and Gupta}{Zhang
  et~al\mbox{.}}{2006}]%
        {zhang2006locating}
\bibfield{author}{\bibinfo{person}{Xiangyu Zhang}, \bibinfo{person}{Neelam
  Gupta}, {and} \bibinfo{person}{Rajiv Gupta}.}
  \bibinfo{year}{2006}\natexlab{}.
\newblock \showarticletitle{Locating faults through automated predicate
  switching}. In \bibinfo{booktitle}{\emph{Proceedings of the 28th
  International Conference on Software Engineering}}.
  \bibinfo{pages}{272--281}.
\newblock


\bibitem[\protect\citeauthoryear{Zhang, Zhai, Ma, and Shen}{Zhang
  et~al\mbox{.}}{2021}]%
        {ZhangAUTOTRAINER}
\bibfield{author}{\bibinfo{person}{Xiaoyu Zhang}, \bibinfo{person}{Juan Zhai},
  \bibinfo{person}{Shiqing Ma}, {and} \bibinfo{person}{Chao Shen}.}
  \bibinfo{year}{2021}\natexlab{}.
\newblock \showarticletitle{AUTOTRAINER: An Automatic DNN Training Problem
  Detection and Repair System}. In \bibinfo{booktitle}{\emph{ICSE'21: The 43nd
  International Conference on Software Engineering}}.
\newblock


\bibitem[\protect\citeauthoryear{Zhang, Ren, Chen, Xiong, Cheung, and
  Xie}{Zhang et~al\mbox{.}}{2020}]%
        {zhang2020detecting}
\bibfield{author}{\bibinfo{person}{Yuhao Zhang}, \bibinfo{person}{Luyao Ren},
  \bibinfo{person}{Liqian Chen}, \bibinfo{person}{Yingfei Xiong},
  \bibinfo{person}{Shing-Chi Cheung}, {and} \bibinfo{person}{Tao Xie}.}
  \bibinfo{year}{2020}\natexlab{}.
\newblock \showarticletitle{Detecting numerical bugs in neural network
  architectures}. In \bibinfo{booktitle}{\emph{Proceedings of the 28th ACM
  Joint Meeting on European Software Engineering Conference and Symposium on
  the Foundations of Software Engineering}}. \bibinfo{pages}{826--837}.
\newblock


\end{thebibliography}


\end{document}